# A new dominance concept and its application to diversity-stability analysis

Zhanshan (Sam) Ma[1*] and Aaron M. Ellison[2]

[1]Computational Biology and Medical Ecology Lab, State Key Laboratory of Genetic Resources and Evolution, Kunming Institute of Zoology, Chinese Academy of Sciences, Kunming, 650223, China

[2]Harvard University, Harvard Forest, 324 North Main Street, Petersham, Massachusetts, MA 01366, USA

*For all correspondence: ma@vandals.uidaho.edu

## Abstract

We introduce a new dominance concept consisting of three new dominance metrics based on Lloyd's (1967) mean crowding index. The new metrics link communities and species, whereas existing ones are applicable only to communities. Our community-level metric is a function of Simpson's diversity index. For species, our metric quantifies the difference between community dominance and the dominance of a virtual community whose mean population size (per species) equals the population size of the focal species. The new metrics have at least two immediate applications: (*i*) acting as proxies for diversity in diversity-stability modeling (*ii*) replacing population abundance in reconstructing species dominance networks. The first application is demonstrated here using data from a longitudinal study of the human vaginal microbiome, and provides new insights relevant for microbial community stability and disease etiology.

## 1. Introduction

The relationship between diversity and stability is of central importance to both theoretical and applied ecology (*e.g.*, Thibaut & Connolly 2013, Wang & Loreau 2016). Sequence-based metagenomics are revealing high levels of microbial diversity, but the stability of microbial communities remains poorly understood (*e.g.*, Lozupone et al. 2012, Oh et al. 2016, Moya & Ferrer 2016). Here, we introduce and develop three new dominance metrics for describing communities of organisms and the species that make up the communities. Our metrics are based on Lloyd's index of mean crowding (Lloyd 1967), and link species and communities in a single framework. We illustrate the use of these new metrics through an exploration of the diversity-stability relationship in the human vaginal microbial community (HVMC), which also serves to provide new insights into the etiology of bacterial vaginosis (BV).

The concept of mean crowding was developed by Lloyd (1967) in his study of population aggregation. We extend this concept from the population (species) scale to encompass assemblages (communities) by defining three new dominance metrics:[1] *community dominance* ($D_c$), *species dominance* ($D_s$), and *species dominance distance* ($D_{sd}$). We use dominance as a proxy for diversity for two reasons. Pragmatically, $D_c$ is a simple linear function of Simpson's diversity index but it is easier to examine dominance and its relationship to stability in the HVMC. More generally, however, the concept of dominance can be applied to both species and communities. For example, we commonly refer to communities with high species diversity, but rarely refer to high-diversity species. However, it often matters greatly which species dominates an assemblage (e.g., Ellison et al. 2005, Valls et al. 2015), and its identity refers to an individual species or population. To the best of our knowledge, there is not an existing index of dominance or diversity can be applied simultaneously to species and assemblages.

In the following sections, we first sketch how to extend Lloyd's (1967) index of mean crowding to both species and community scales (complete technical details and mathematical proofs are provided in Supplemental Online Material). We then develop a novel phenomenological modeling approach to dominance (diversity)–stability relationships, focusing on five linear and non-linear diversity-stability models (linear, logistic, sine-logistic, linear-quadratic, and quadratic-quadratic). This approach reveals three fundamental components of the diversity-stability relationship: dominance-dependent stability (DDS), dominance-inversely-dependent stability (DID), and dominance-independent stability (DIS). Our modeling approach distinguishes between stability and resilience; the latter is defined as the derivative of the former, and characterizes the rate of changes in stability, analogously to the relationship between acceleration and velocity. Last, the models illustrate the (in)stability of the community equilibrium in HVMC.

---

[1] We note that in the community ecology literature, *index, measures,* and *metrics* have been used interchangeably. Here, we use *metric* in a general sense (as a "type"), as opposed to *index*, which is an "instance" of a metric.

## 2. Materials and Methods

### 2.1. Community and species dominance based on mean crowding

Lloyd (1967) defined aggregation or dispersion of an animal population as "the mean number per individual of other individuals in the same quadrat" (Lloyd 1967) or "the average number of other individuals per quadrat, per individual" (Lloyd 1986). Mathematically, mean crowding ($m*$) is calculated as:

$$m* = m + \frac{\sigma^2}{m} - 1 \qquad (1)$$

where $m$ is population density (abundance) and $\sigma^2$ is its corresponding variance. Lloyd (1967) suggested mean crowding would be particularly suitable for measuring free-moving animals in a relatively continuous habitat. Therefore, the concept should be applicable to free-moving bacterial species in the largely continuous habitat of the HVMC.

Measures of aggregation, dispersion, patchiness, heterogeneity, skewness, evenness, and dominance often are used to characterize the abundance distribution of biological species; their temporal variability often is associated with the (in)stability of populations. Because aggregation can be considered as inversely related to evenness, extending mean crowding to species dominance (*i.e.*, unevenness) should be straightforward.

We start by defining mean crowding of a community as:

$$m_c* = m_c + \frac{\sigma_c^2}{m_c} - 1 \qquad (2)$$

where $m_c$ = the mean population abundance (size) per species of all species in the community and $\sigma_c^2$ is its variance. Note that the $m_c$ is computed across species, not samples. By analogy with $m^*$, $m_c^*$ is a measure of community unevenness or dominance (aggregation per species), and we interpret $m_c^*$ as the average number of other individuals per species, per individual. Essentially, we treat "species" as a virtual quadrat (*sensu* Lloyd 1967); with quadrat being the sampling unit for $m^*$ and species being the sampling unit for $m_c^*$.

Since there are many species in the community, it makes sense to divide $m_c^*$ by the mean population size (abundance) per species, $m_c$.

$$D_c = \frac{m_c^*}{m_c} = 1 + \frac{\sigma_c^2}{m_c^2} - \frac{1}{m_c}, \tag{3}$$

which is the direct counterpart of population-level patchiness or heterogeneity of population distribution (Lloyd 1967; Iwao 1968). For communities, we interpret $D_c$ as *community dominance* (unevenness) in terms of the deviation from the average species. Analogously with population aggregation, $D_c$ also can be interpreted as the "center-of-gravity" of a community, measuring how crowded the individuals of the average species are crowded by the individuals of its neighbor species, and it is essentially the counterpart of population aggregation (unevenness) in population ecology. We also observe that $D_c$ is linearly related to Simpson's index $D$:

$$D_c = nD - n \Big/ \sum_{i=1}^{n} m_{s_i} \tag{4}$$

where $n$ is the number of species in the community, and $m_{s_i}$ is the mean abundance (size) of the $i^{th}$ species (see mathematical proof in Supplemental Online Material). Besides the mathematical proof, we also examine the relationship between $D_c$ and more familiar measures of community dominance or diversity, *viz.* Shannon-Weiner $H'$, Simpson's index $D$, and the Berger-Parker index of dominance (Berger & Parker 1970) as a "sanity check" on our analogy. We fit a simple linear model between $D_c$ and the existing dominance indexes (Table 1). In particular, the relationship with Simpson's index, as expected, exact ($r$=1).

We next use $D_c$ to define a dominance index for each species in the community. We define the *species dominance distance* as:

$$D_{sd} = \frac{m_c^*}{m_s} = \frac{m_c}{m_s} + \frac{\sigma_c^2}{m_c m_s} - \frac{1}{m_s}, \tag{5}$$

which defines a dominance index that "distributes" the mean community crowdedness ($m_c^*$) over a specific species ($m_s$) rather than over an average species ($m_c$). $D_{sd}$ ranges from 0 to $+\infty$. Although the behavior of $D_{sd}$ counterintuitive—dominant species may have large values of $m_s$ and hence small values of $D_{sd}$—its interpretation is intuitive if we imagine community as a sphere with a center of gravity = $D_c$ with dominant species (in terms of abundances) closer to its center than rare ("satellite") species. This is also why we call $D_{sd}$ *species dominance distance* to differentiate it from *species dominance*, which we define next.

Last, we define *species dominance* (*index*) (range = $-\infty$ to $+\infty$) as the difference between community dominance ($D_c$) and species dominance distance ($D_{sd}$):

$$D_s = D_c - D_{sd} = \frac{m_c^*}{m_c} - \frac{m_c^*}{m_s} = 1 - \frac{m_c}{m_x} + \frac{\sigma_c^2}{m_c^2} - \frac{\sigma_c^2}{m_c m_s} + \frac{m_c - m_s}{m_c m_s}. \tag{6}$$

Dominant species have larger values of $D_s$ than other species. It is noted that we had two considerations to define species dominance. First, as demonstrated elsewhere (Ma & Ellison 2017), *species dominance* (index) is more suitable than *species dominance distance* for conducting species dominance network analysis. Second, we adopted the difference between community dominance and species dominance distance, rather than possibly the inverse of species dominance distance, to avoid non-linear transformation.

## 2.2. Phenomenological modeling of community dominance (diversity), and stability

We start by defining *community dominance stability* (shortened to *community stability* hereafter) as:

$$S_c(t) = \frac{D_c(t+1) - D_c(t)}{D_c(t)}. \tag{7}$$

$S_c(t)$ measures the change rate of community dominance (for other definitions of ecological stability, see Grimm & Wissel 1997). Similarly, we define *population dominance stability* (hereafter, *population stability*) as the change rate of species dominance over time,

$$S_s(t) = \frac{D_s(t+1) - D_s(t)}{D_s(t)}, \tag{8}$$

noting that there would be a separate measure of stability for each species in the community. Finally, we assume that community dynamics can be modeled by a general (set of) differential equation(s) such as:

$$S(t) = \frac{dD(t)}{D(t)dt} = f[D(t), Z] \tag{9}$$

where $D(t)$ is one of the dominance metrics at time $t$, $S(t)$ is the parallel stability metric and time $t$, and $Z$ is an optional vector of covariates

Since we do not know the categorical form of function $f$ in Eqn. 8, the modeling strategy we take is data-driven and phenomenological. Through trial-and-error and exploratory curve-fittings, we found that the following five models are promising in describing the community dominance dynamics: a two-parameter linear model $S_c(t) = a + bD_c(t)$; a five-parameter linear-quadratic (L-Q) model $S_c(t) = a + bD_c(t) + cD_c^2(t) + [D_c(t) - d]Sign(D_c(t) - d)[c(D_c(t) + d) + e]$;

a six-parameter quadratic-quadratic (Q-Q) model $S_c = a + bD_c + cD_c^2 + (D_c - d)Sign(D_c - d)[e(D_c + d) + f)]$ ; a three-parameter logistic $S_c(t) = \dfrac{K}{1 + a * \exp[-rD_c(t)]}$ ; and a three-parameter periodic logistic-sine model $S_c(t) = \dfrac{K}{1 + a * \exp[-rD_c(t)]} \sin\left(\dfrac{D_c(t)}{\pi}\right)$ . A detailed discussion of these five models and the derivations of their parameters are given in the Supplemental Online Material.

For each of the stability models listed above, there is a corresponding dominance dynamics model. For community dominance, this would be:

$$D_c(t+1) = D_c(t)[1 + S_c(t)] \qquad (10)$$

and the corresponding model for dominance dynamics with the three-parameter logistic stability model would be, for example:

$$D_c(t+1) = D_c(t)\{1 + K / [1 + a \exp(-rD_c(t)]\} . \qquad (11)$$

Similar to density-dependence models for population regulation (*e.g.*, Berryman 1998; Kot 2001; Pastor 2008), the generalized stability model (Eqns. 7-9) may display three types of local behavior: (*i*) dominance-dependent stability (DDS), in which stability increases with dominance; (*ii*) dominance-inversely-dependent stability (DID), in which stability decreases with dominance; and (*iii*) dominance-independent stability (DIS), in which stability does not change with dominance level. That is: $S(t) \propto kD(t)$, with $k < 0$, $k > 0$, $k = 0$, corresponding respectively to cases (*i*), (*ii*) and (*iii*).

In practice, except for the simple two-parameter linear model (where the parameter *b* is equivalent to *k* in the generalized stability model), we may not be able to determine the value of *k*. However, the nonlinear models we considered are still simple enough that we can evaluate the piece-wise relationship between dominance and stability. We also note that the three-parameter logistic and two-parameter linear model may only capture one of the three dominance-dependence behaviors in a specific model, but the other three models are more flexible and may capture all three behaviors in a single model. The logistic-sine model also can capture periodic fluctuation of the three types of dependence relationships.

Since our dominance concept (metrics) can act as appropriate proxies for diversity, the above-described modeling approach equivalently offers an equally powerful method for modeling classic diversity-stability relationships.

## 2.3. The HVMC dataset

We compare our dominance metrics to three other diversity indices (the "sanity check" described above) and illustrate the phenomenological modeling and selection among the five candidate dominance-stability model using a "32-healthy cohort dataset" from a longitudinal study of 32 healthy women conducted at Johns Hopkins University School of Medicine, Baltimore, MD, from 2006 to 2007 (Gajer *et al.* 2012). The participants were advised to self-collect mid-vaginal swabs and vaginal smears twice weekly for 16 weeks. The extraction of genomic DNA from frozen vaginal swabs, PCR amplification and sequencing of the V1-V2 region of bacterial 16S rRNA genes were described in details in Gajer *et al.* (2012). The archive of sequence data is described in Gajer et al. (2012). The QIIME (Caporaso *et al.* 2010), UCLUST (Edgar *et al.* 2010), UCHIME (Edgar *et al.* 2011), RDP Naïve Bayesian Classifier (Wang 2007) and speciateIT (speciateIT.sourceforge.net) were utilized to obtain the OTU table (including sequence read counts and relative abundances of the taxonomic assignments at the bacterial species level of 97% similarity, see Gajer et al. 2012 for detailed descriptions).

# 3. Results and Discussion

## 3.1. Computation and interpretation of new dominance metrics, and comparisons between dominance metrics and diversity indices

Figure 1 (drawn based on the linear regression parameters in Suppl. Table S1) compares the new community dominance ($D_c$) metric with three other diversity indexes. As $D_c$ is a linear transformation of Simpson's $D$, those two are perfectly correlated ($R$=1). Correlations between $D_c$ and both Shannon indexes and the Berger-Parker diversity index exceeded 0.95 among all but two of the HVMC swabs ($r$ for those two outliers equaled 0.94 and 0.75), and were statistically significant in all cases ($P < 0.001$). The overall correlation between $D_c$ and Simpson's measure of evenness ($D/S$) averaged 0.61 (median $r = 0.69$; Fig. 1) and again was statistically significant ($P < 0.001$). We conclude that $D_c$ is quantitatively comparable to existing measures of diversity.

Suggested Location for Figure 1

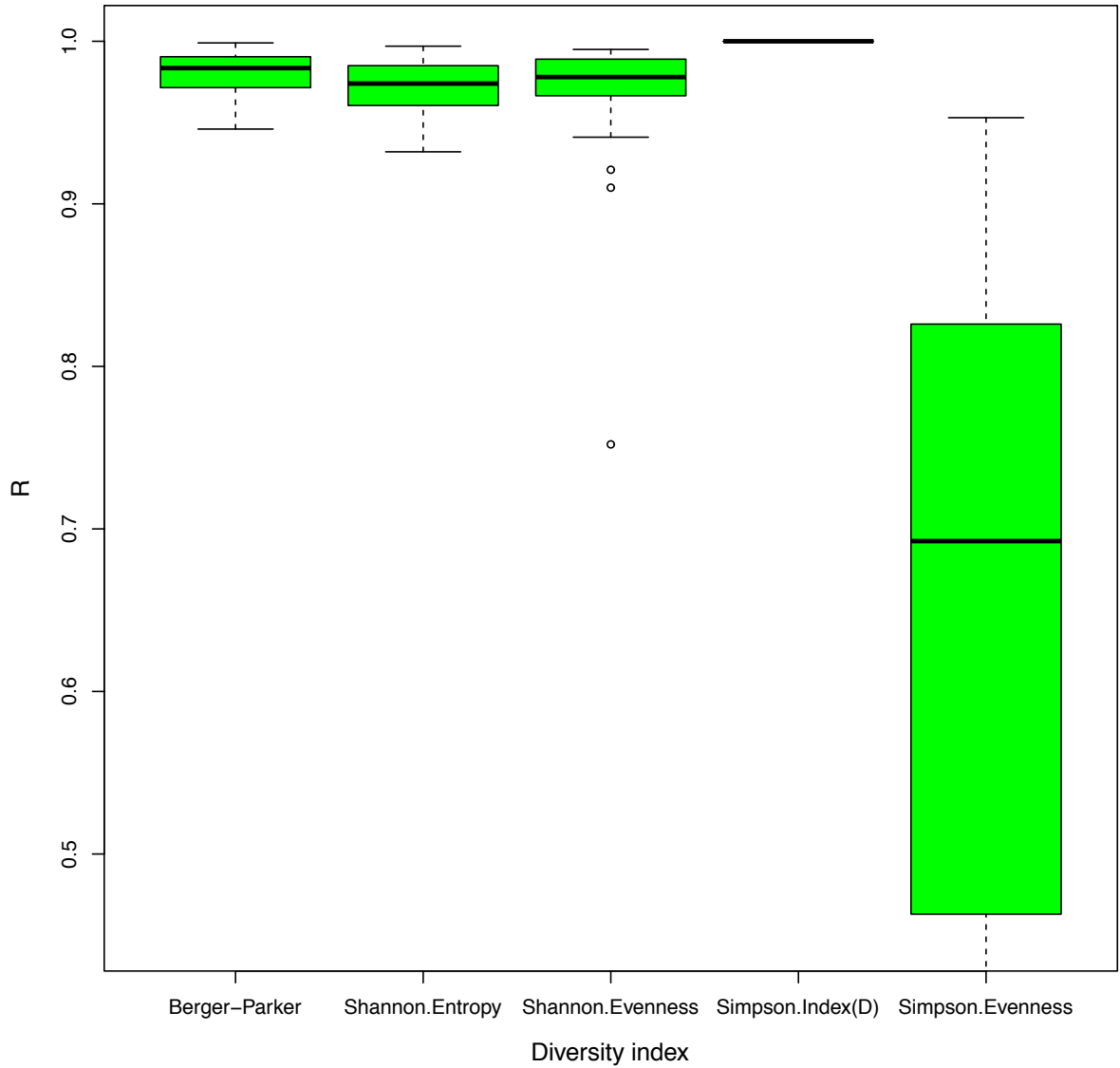

**Figure 1**. A graph showing the perfect linear correlation ($R$=1) between the new dominance metric and Simpson's index ($D$) as well as the statistically significant linear relationships between the new metric and other four existing diversity indexes. The perfect correlation is expected from Eq. (4) and its mathematical proof is presented in Online Supporting Information. The slightly poor correlation with Simpson's evenness is also discussed in Online Supporting Information.

Why introduce another diversity (dominance) metric? We assert that the utility of $D_c$ is its straightforward extension to, and interpretation at, the species level. This extension ($D_s$) allows us to quantitatively identify which species dominates a community and to what extend it dominates the community. An example is shown in Table 1 for seven microbial species from one of the subjects in the 32-healthy cohort HVMC dataset (the results for all species of the subject (#400) are given in Supplementary Online Material).

Suggested Location for Table 1

Note that theoretically $D_{sd} \in (0, +\infty)$, $D_s \in (-\infty, +\infty)$. Because $D_{sd} = +\infty$ and $D_s = -\infty$ when the abundance of the focal species = 0, we replace $D_s = -\infty$ in a particular sample with its lowest value in all the time-series samples of that subject. We note that at the most extreme value (abundance = 0, $D_s = -\infty$) that the distance $D_{sd} \rightarrow +\infty$ and the species becomes "disconnected" from the community, corresponding to the temporarily or local extinction of the species. Of course, it is possible to artificially convert the metric value into a small range such as [0, 1], but we do not see any need or benefit from it. In fact, the capability to represent discontinuous points or local (temporary) extinctions should be an advantage, especially at the species level, because temporary disappearance or local extinction of bacterial species is frequently observed in the HVMC (*e.g.,* Gajer et al. 2012). Obviously, the *dominance* rank and *abundance* rank can be very different because the most abundant species are not necessarily the most dominant species, and the least abundant species are not necessarily the least dominant, and *vice versa*.

We also observe that less common (rare species) may have even larger fluctuations in their $D_s$ values. Time-series of $D_{sd}$ and $D_s$ effectively illustrate this phenomenon (Figure 2), and we think that such illustrations (*e.g.,* as exploratory data analysis) could help identify potential associations between rare microbes and pathological changes (*e.g.,* opportunistic pathogens should be rare, at least initially) such as the occurrence of BV. Indeed, this was one of the major motivations for our development of dominance metrics that could be used at both community and species levels. Figure 2 shows the community dominance as well as the species dominance of 7 selected species including top three most abundant species, two moderately abundant species, and two least abundant (rare) species. Figure 3 exhibits community dominance and species dominance of top 3 most abundant species. Both Figures 2 and 3 were drawn based on the time-series data (29 sampling points) of subject #400, which are provided in Table 1 and Table S7. We also found that a graph such as Figure 3 is especially effective when drawn in polar coordinates. Figure 3 reveals that community dominance seems to be "controlled" by one of the top 3 species (two circles representing the community dominance and the "master" species overlapped), but the "control" is dynamic and may be transferred from one species to another. Nevertheless, we realized that such kind of traditional graphic representation is still rather limited in dealing with complex issues such as detecting the potential "culprits" or "control mechanism" in BV etiology, even with our new dominance metrics and with the relatively powerful polar coordinate system. Instead, we found that species co-dominance analysis supported by our new metrics is much more effective (Ma & Ellison 2017). For example, it was Figure 3 that inspired us to search for species trio motif in the species co-dominance network, which turned out to nicely explain the possible control mechanisms of community dominance (Ma & Ellison 2017).

Suggested Location for Figure 2
Suggested Location for Figure 3

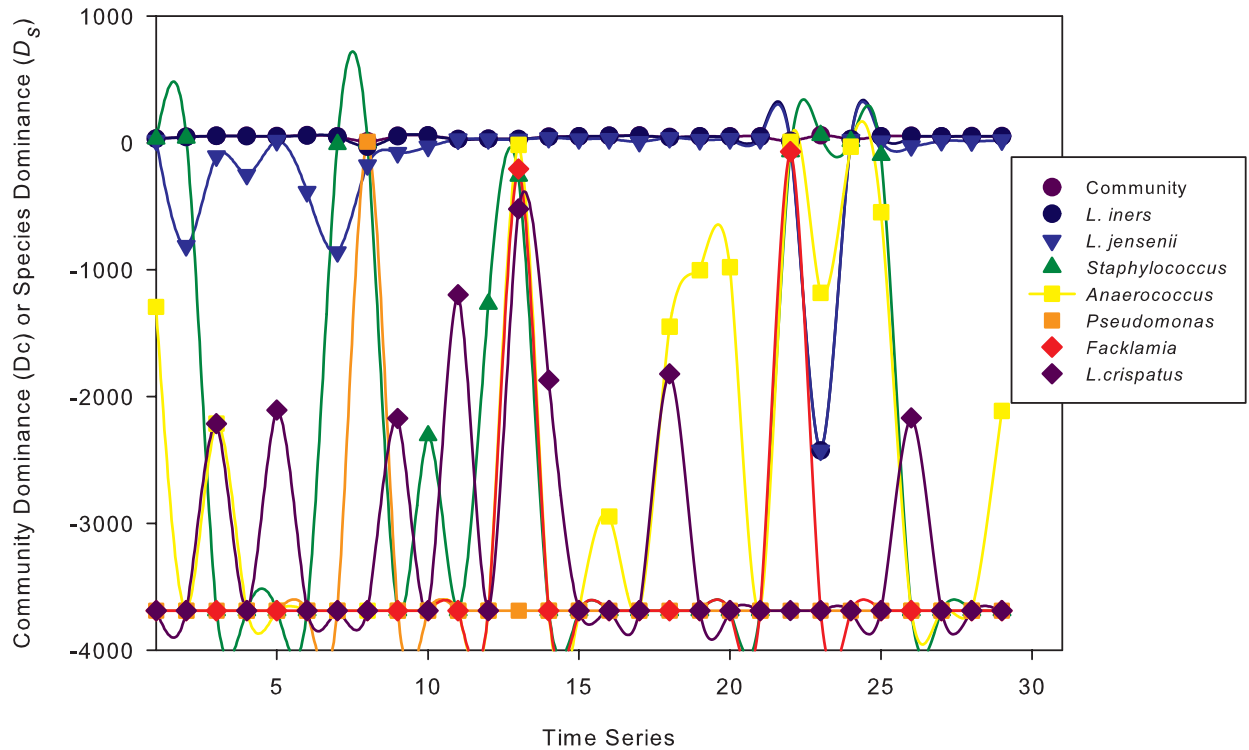

**Figure 2**. Community dominance ($D_c$) and species dominance ($D_s$) metrics of seven representative species selected from the time-series data of Subject#400, including the three most abundant species (*L. iners, L. jensenii, & Staphylococcus*), two moderately abundant species (*Anaerococcus & Pseudomonas*) and the two least abundant (rare) species (*Facklamia & L. crispatus*).

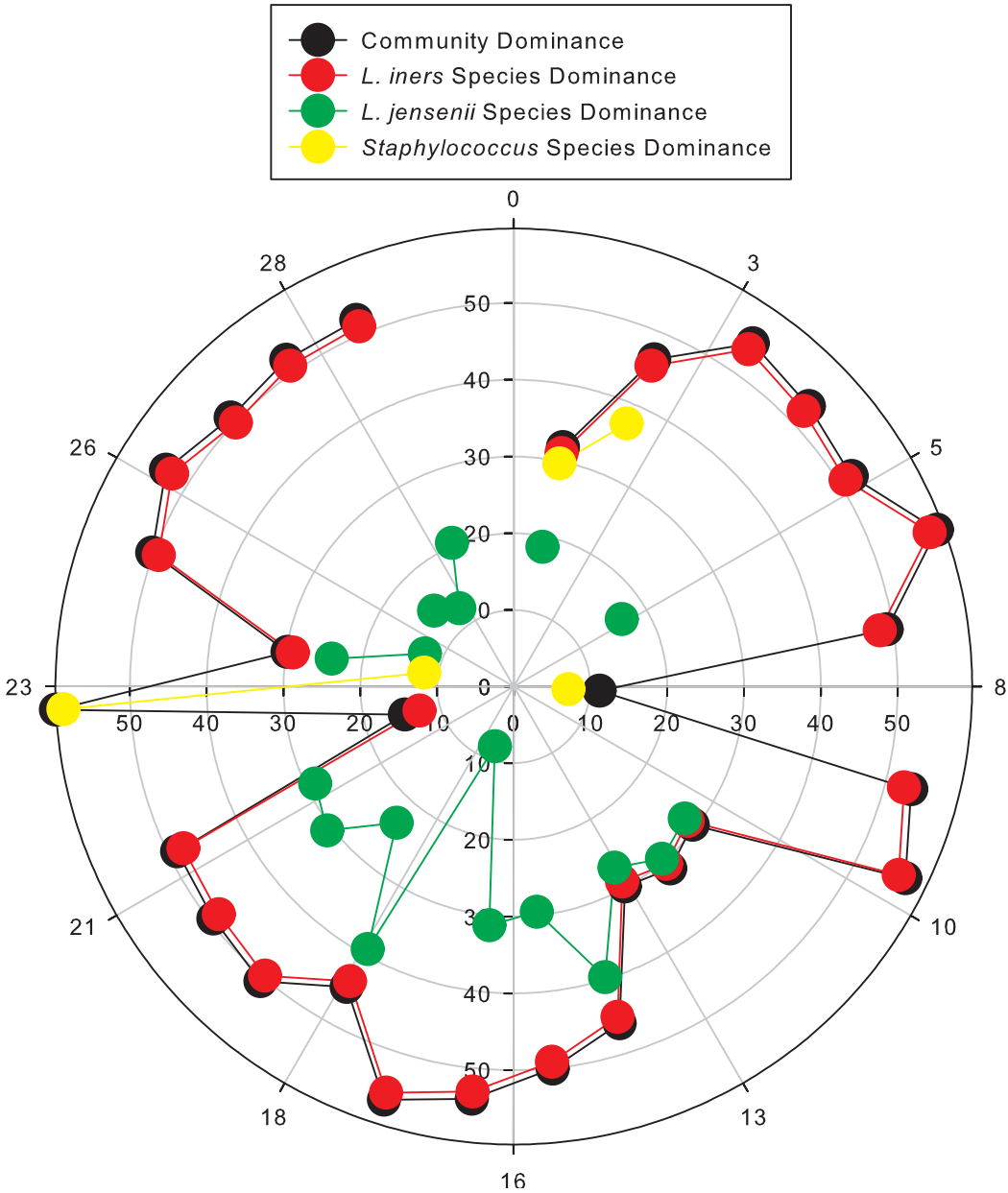

**Figure 3**. A polar coordinate graph showing community dominance and the species dominance metrics of the three most abundant species in the vaginal microbial community of Subject#400

It could be argued that other diversity or evenness indices could be extended similarly, but it would not be easy. For example, the Berger-Parker (1970) diversity index yields the same value if the most abundant species in two communities are equally abundant in each community, regardless of whether they are the same species. However, two communities may be dominated by different species with equal abundances.

## 3.2. Phenomenological modeling of community dominance and stability

Our focus here is on illustrating qualitative patterns of stability and its relationship with measures of dominance as modeled with linear, linear-quadratic (L-Q), quadratic-quadratic (Q-Q), logistic, and logistic-sine models. For each model, we examined whether it revealed biologically interpretable patterns and how well it fit the 32-healthy cohort HVMC data. Our modeling strategy is a compromise between realism and simplicity, and includes biological interpretations of model parameters, statistical tests (coefficient of determination $r^2$ and standard errors of parameters), and an appeal to parsimony. Details of model selection are discussed further in Supplementary Online Material.

The most appropriate model for each subject is given in Table 2, which were selected from Suppl. Tables (S2-S6) based on the above-mentioned modeling strategy. There is a significant difference between the *logistic model* and *linear model*, and the other three models. Both the logistic model and the linear model can capture only one kind of the diversity-stability mechanisms with a single model, depending on the sign of the parameter $b$ (linear model) or $r$ (logistic model). In contrast, the other three models, *i.e.*, logistic-sine, L-Q, and Q-Q models may capture all three diversity-stability mechanisms, *i.e.*, DDS, DIS, and DID, simultaneously with a single model. That is, the same community may exhibit three diversity-stability mechanisms alternately. The above-described difference between two categories of the models (linear & logistic models *vs*. logistic-sine, L-Q and Q-Q) is obvious in Table 2. The latter category of models can describe more complex diversity-stability mechanisms. For example, alternating DDS, DID and DIS with multiple equilibriums in the case of logistic-sine model.

<div align="center">Suggested Location for Table 2</div>

The parameter ($r$) of the two logistic models and the slope ($b$) of the linear models are negative for all subjects; hence, there is no need to note the sign of the model parameters in Table 2. For the HVMC data, the dominance-stability relationships modeled by the logistic and linear models are dominance-dependent, *i.e.*, the higher the dominance, the more stable the community. Equivalently, when diversity is high (*i.e.*, dominance is low), stability declines. However, we emphasize that not all of the HVMCs exhibit dominance-dependent stability (DDS), as suggested by the other three models (*i.e.*, Logistic-sine, L-Q and Q-Q models). In both the L-Q and Q-Q models, the combinations of parameter values of $b_1$ & $c_2$ or $c_1$ & $c_2$ can be used to determine the types of dominance-stability relationship and equilibria; hence the signs of these parameters are noted in Table 2 to facilitate the determination of dominance-stability relationships.

Additional examination of the slope ($b$) of the linear models [Suppl. Table (S4)], however, suggests additional nuances in the relationship between dominance (diversity) and stability in the HVMC. The slope $b$ is the derivative of the linear stability function, *i.e.*, the rate of change of stability with respect to dominance. This slope is a measure of resilience: the speed at which a community returns to local equilibrium after perturbation.

Until recently, the prevailing opinion was that more diverse (higher diversity) HVMCs are less stable and prone to BV. Although Ma et al. (2012) rightly rejected the opinion by citing counter examples, they did not present a mechanistic explanation of this important syndrome in BV etiology. Indeed, the rejection of this opinion can be counted as one of the most significant findings obtained from the metagenomic studies of the HVMC and BV in the last few years (Ravel et al. 2011, Gajer et al. 2012). Our identification of dominance-dependent stability mechanism as displayed by the standard logistic and the linear model (see Suppl. Table S2 & S4) appears to support the previous prevalent opinion because dominance-dependence mechanism predicts that lower dominance (higher diversity) corresponds to lower stability. However, if we carefully analyze the implication of the slope ($b$) of the linear models, a counter-argument to the above-mentioned prevalent opinion emerges. The slope $b$ is the *derivative* of the linear stability function, *i.e.*, the change rate of stability with respect to dominance. A community that has a steeper slope ($b$) should be easier to stabilize with the same units of dominance increase than a community with a less steep slope.

The range of the difference in slope ($b$) among communities in Suppl. Table S4 is rather wide, exceeding 15 times (smallest $b=-0.123$ for subject #412, and largest $b=-0.008$ for subject#443). This suggests that a *diverse community*—the community that usually lacks apparent dominant species—is not necessarily inherently unstable because it can be quicker in stabilizing itself than a counterpart that is with highly dominant species.

The above apparent contradiction can be easily resolved by a careful distinction between the community *stability* and *resilience*. There are numerous definitions of stability (*e.g.*, Grimm & Wissel 1997), and resilience is often treated as one component or dimension of stability (Ma 2012). We defined community stability as $S_c(t)$ [Eq. (7)], but in the case of linear model, the slope ($b$) is actually a representative of the community *resilience*, which is often defined as the 'speed' at which community returns to local equilibrium after perturbation. It should be noted that in our definition of community stability, time ($t$) is implicitly included in the stability function [Eq. (7)-(9)]; therefore, the slope ($b$) of the linear model is indeed a measure of resilience because $b$ is rightly the *derivative* of the linear model.

By distinguishing community stability from resilience in the context of the linear stability function, we can draw the following insights from Suppl. Table S4: The dominance-dependent stability mechanism suggests that high diversity (low dominance) community can be less stable than low diversity (high dominance) community, but the former, if its slope is steeper than the slope of the latter, may have higher resilience than the latter. This distinction resolves the apparent paradox regarding the diversity-stability relationship in the case of HVMC, and presents a more comprehensive, cohesive and quantitative argument to support Ma et al.'s (2012)'s rejection of the previously prevalent opinion on BV.

In summary, when discussing community stability, it is critical to distinguish between community *stability* and *resilience*. Both stability and resilience are needed to accurately describe the diversity-stability relationship in HVMC. A high diversity community may have lower stability in terms of the magnitude of community dominance change, as suggested by the prevalent opinion on the stability of HVMC, but the community may still be resilient. In other words, a high diversity community may possess more efficient/effective mechanism to stabilize itself. This of course, is not difficult to explain with now well-known diversity-stability paradigm theory in macro ecology. That is, high diversity community may have higher connectivity that leads to more efficient mechanism in stabilizing the community. Therefore, it is clear that high diversity community, which usually lacks apparently dominant species, is not necessarily unstable inherently because it can be more resilient.

## Acknowledgements

We appreciate Prof. Larry Forney, University of Idaho, and Jacques Ravel, University of Maryland for discussion of BV at various occasions. We are also indebted to DD Ye, LW Li, and P Dong, Chinese Academy of Sciences for verifying the computational results.

**Conflict of Interest**: The authors declare no conflict of interests.

**Supplementary information** is available at the ISMEJ's website www.nature.com/ismej

**Table 1**. The *community dominance metric* ($D_c$), *species dominance distances* ($D_{sd}$) and *species dominance metric* ($D_s$) of seven species selected from the 29 longitudinal samples of Subject #400 [*, **]

| Sample ID | Community Dominance $D_c$ | OTU#1 $D_{sd}$ | $D_s$ | OTU#8 $D_{sd}$ | $D_s$ | OTU#28 $D_{sd}$ | $D_s$ | OTU#11 $D_{sd}$ | $D_s$ | OTU#115 $D_{sd}$ | $D_s$ | OTU#57 $D_{sd}$ | $D_s$ | OTU#2 $D_{sd}$ | $D_s$ |
|---|---|---|---|---|---|---|---|---|---|---|---|---|---|---|---|
| 400_010106 | 31.824 | 0.774 | 31.050 | 13.259 | 18.566 | 2.175 | 29.649 | 1325.881 | -1294.056 | ∞ | -3688.829 | ∞ | -3688.829 | ∞ | -3688.829 |
| 400_010506 | 46.355 | 0.884 | 45.471 | 858.504 | -812.149 | 9.090 | 37.265 | ∞ | -3688.829 | ∞ | -3688.829 | ∞ | -3688.829 | ∞ | -3688.829 |
| 400_010806 | 54.460 | 0.953 | 53.507 | 159.273 | -104.812 | ∞ | -3688.829 | 2269.639 | -2215.178 | ∞ | -3688.829 | ∞ | -3688.829 | 2269.639 | -2215.178 |
| 400_011206 | 53.032 | 0.941 | 52.090 | 304.780 | -251.748 | ∞ | -3688.829 | ∞ | -3688.829 | ∞ | -3688.829 | ∞ | -3688.829 | ∞ | -3688.829 |
| 400_011906 | 51.848 | 0.931 | 50.918 | 35.271 | 16.577 | ∞ | -3688.829 | ∞ | -3688.829 | ∞ | -3688.829 | ∞ | -3688.829 | 2160.351 | -2108.503 |
| 400_012206 | 58.738 | 0.989 | 57.749 | 444.944 | -386.205 | ∞ | -3688.829 | ∞ | -3688.829 | ∞ | -3688.829 | ∞ | -3688.829 | ∞ | -3688.829 |
| 400_012606 | 49.126 | 0.906 | 48.220 | 909.746 | -860.619 | 61.102 | -11.976 | ∞ | -3688.829 | ∞ | -3688.829 | ∞ | -3688.829 | ∞ | -3688.829 |
| 400_012906 | 11.194 | 42.400 | -31.206 | 186.560 | -175.366 | 4.056 | 7.138 | ∞ | -3688.829 | 3.969 | 7.224 | ∞ | -3688.829 | ∞ | -3688.829 |
| 400_020206 | 53.449 | 0.945 | 52.504 | 132.944 | -79.495 | ∞ | -3688.829 | ∞ | -3688.829 | ∞ | -3688.829 | ∞ | -3688.829 | 2226.807 | -2173.359 |
| 400_020506 | 56.791 | 0.973 | 55.818 | 86.030 | -29.239 | 2365.823 | -2309.032 | ∞ | -3688.829 | ∞ | -3688.829 | ∞ | -3688.829 | ∞ | -3688.829 |
| 400_020906 | 29.476 | 0.825 | 28.651 | 1.329 | 28.147 | ∞ | -3688.829 | ∞ | -3688.829 | ∞ | -3688.829 | ∞ | -3688.829 | 1227.941 | -1198.465 |
| 400_021206 | 31.273 | 0.803 | 30.470 | 1.655 | 29.618 | 1303.154 | -1271.881 | ∞ | -3688.829 | ∞ | -3688.829 | ∞ | -3688.829 | ∞ | -3688.829 |
| 400_021606 | 29.827 | 0.730 | 29.097 | 2.816 | 27.011 | 292.364 | -262.537 | 46.889 | -17.061 | ∞ | -3688.829 | 236.676 | -206.849 | 552.243 | -522.416 |
| 400_021906 | 46.040 | 0.884 | 45.156 | 6.310 | 39.731 | ∞ | -3688.829 | ∞ | -3688.829 | ∞ | -3688.829 | ∞ | -3688.829 | 1918.155 | -1872.114 |
| 400_022606 | 50.081 | 0.915 | 49.166 | 20.559 | 29.523 | ∞ | -3688.829 | ∞ | -3688.829 | ∞ | -3688.829 | ∞ | -3688.829 | ∞ | -3688.829 |
| 400_030206 | 54.013 | 0.949 | 53.063 | 22.730 | 31.282 | ∞ | -3688.829 | 3000.408 | -2946.395 | ∞ | -3688.829 | ∞ | -3688.829 | ∞ | -3688.829 |
| 400_030506 | 56.446 | 0.970 | 55.476 | 48.249 | 8.197 | ∞ | -3688.829 | ∞ | -3688.829 | ∞ | -3688.829 | ∞ | -3688.829 | ∞ | -3688.829 |
| 400_030906 | 44.804 | 0.874 | 43.929 | 5.644 | 39.159 | ∞ | -3688.829 | 1493.453 | -1448.649 | ∞ | -3688.829 | ∞ | -3688.829 | 1866.816 | -1822.013 |
| 400_031206 | 50.630 | 0.920 | 49.710 | 27.218 | 23.412 | ∞ | -3688.829 | 1054.695 | -1004.064 | ∞ | -3688.829 | ∞ | -3688.829 | ∞ | -3688.829 |
| 400_031606 | 49.449 | 0.910 | 48.540 | 18.772 | 30.678 | ∞ | -3688.829 | 1030.093 | -980.644 | ∞ | -3688.829 | ∞ | -3688.829 | ∞ | -3688.829 |
| 400_031906 | 48.820 | 0.904 | 47.916 | 20.039 | 28.781 | ∞ | -3688.829 | ∞ | -3688.829 | ∞ | -3688.829 | ∞ | -3688.829 | ∞ | -3688.829 |
| 400_032306 | 14.629 | 1.961 | 12.668 | 23.229 | -8.601 | 84.107 | -69.478 | 6.222 | 8.407 | ∞ | -3688.829 | 84.107 | -69.478 | ∞ | -3688.829 |
| 400_032606 | 59.687 | 2486.726 | -2427.038 | 2486.726 | -2427.038 | 0.997 | 58.690 | 1243.363 | -1183.676 | ∞ | -3688.829 | ∞ | -3688.829 | ∞ | -3688.829 |
| 400_033006 | 29.794 | 0.714 | 29.081 | 5.830 | 23.964 | 17.997 | 11.797 | 59.846 | -30.052 | ∞ | -3688.829 | ∞ | -3688.829 | ∞ | -3688.829 |
| 400_040206 | 50.181 | 0.916 | 49.265 | 37.855 | 12.326 | 149.392 | -99.211 | 597.567 | -547.387 | ∞ | -3688.829 | ∞ | -3688.829 | ∞ | -3688.829 |
| 400_040606 | 53.354 | 0.944 | 52.410 | 78.693 | -25.339 | ∞ | -3688.829 | ∞ | -3688.829 | ∞ | -3688.829 | ∞ | -3688.829 | 2223.079 | -2169.725 |
| 400_040906 | 50.857 | 0.922 | 49.935 | 36.539 | 14.318 | ∞ | -3688.829 | ∞ | -3688.829 | ∞ | -3688.829 | ∞ | -3688.829 | ∞ | -3688.829 |
| 400_041306 | 51.842 | 0.931 | 50.912 | 39.450 | 12.392 | ∞ | -3688.829 | ∞ | -3688.829 | ∞ | -3688.829 | ∞ | -3688.829 | ∞ | -3688.829 |
| 400_041606 | 51.965 | 0.931 | 51.033 | 31.615 | 20.350 | ∞ | -3688.829 | 2165.637 | -2113.672 | ∞ | -3688.829 | ∞ | -3688.829 | ∞ | -3688.829 |

*Top three most abundant species: OTU#1=*L. iners*, OTU#8=*L. jensenii*, OTU#28=*Staphylococcus*. Two moderate abundant species are: OTU#11=*Anaerococcus,* OTU#115=*Pseudomonas.* Two least abundant species (but excluded species with total reads<10) are: OTU#57=*Facklamia*, OTU#2=*L. crispatus.*

** Note that theoretically $D_{sd} \in (0, +\infty)$, $D_s \in (-\infty, +\infty)$. When population abundance=0, $D_{sd}=\infty$ and $D_s=-\infty$. In practice, for each subject, we replace the $(-\infty)$ with the *species dominance* value of the least dominant member in all time-series samples of the subject. Obviously, dominance rank and abundance rank can be very different because the most abundant species are not necessarily the most dominant species, and least abundant species are not necessarily least dominant.

**Table 2**. A summary of the five models including logistic, logistic-sine, linear, L-Q, and Q-Q models fitted to 32-healthy cohort, summarized from Suppl. Tables (2)-(6)

| Subject ID | Model (R for linear or $R^2$ for others) | Dominance-Stability Relationship |
|---|---|---|
| #400 | Logistic (0.91) | DDS with an asymptotic equilibrium line when $D_c \rightarrow \infty$ |
| #412 | Logistic (0.97) | |
| #439 | Logistic (0.87) | |
| #420 | Logistic-Sine (0.71) | DDS and DIS alternate periodically |
| #431 | Logistic-Sine (0.36) | |
| #402 | L-Q [0.68; $b_1>0$, $c_2>0$] | DIS followed by DDS, a possible stable equilibrium and DIS |
| #416 | L-Q [0.74; $b_1>0$, $c_2>0$] | |
| #429 | L-Q [0.67; $b_1>0$, $c_2>0$] | |
| #408 | L-Q [0.85; $b_1<0$, $c_2>0$] | DDS followed by a possible equilibrium and DIS |
| #445 | L-Q [0.66; $b_1<0$, $c_2>0$] | |
| #411 | L-Q [0.68; $b_1<0$, $c_2<0$] | DDS followed by possibly two equilibriums and DDS |
| #423 | L-Q [0.77; $b_1<0$, $c_2<0$] | |
| #435 | L-Q [0.73; $b_1<0$, $c_2<0$] | |
| #436 | L-Q [0.47; $b_1<0$, $c_2<0$] | |
| #437 | L-Q [0.54; $b_1<0$, $c_2<0$] | |
| #415 | Q-Q [0.73; $c_1<0$, $c_2>0$] | DDS and DIS alternate, two parabolas connected at $D_c=d\approx18$, with a possible stable equilibrium. |
| #418 | Q-Q [0.86; $c_1<0$, $c_2<0$] | DDS and DIS alternate, two parabolas connected at $D_c=d\approx43$, stability of equilibriums is uncertain. |
| #446 | Q-Q [0.59; $c_1<0$, $c_2<0$] | DDS and DIS alternate, two parabolas connected at $D_c=d\approx43$, stability of equilibriums is uncertain. |
| #401 | Linear (R=0.53) | Globally DDS, but the mechanism may be complex locally. |
| #403 | Linear (R=0.80) | |
| #404 | Linear (R=0.37) | |
| #405 | Linear (R=0.84) | |
| #406 | Linear (R=0.52) | |
| #407 | Linear (R=0.48) | |
| #410 | Linear (R=0.49) | |
| #413 | Linear (R=0.45) | |
| #414 | Linear (R=0.57) | |
| #424 | Linear (R=0.68) | |
| #430 | Linear (R=0.45) | |
| #432 | Linear (R=0.63) | |
| #443 | Linear (R=0.43) | |
| #444 | Linear (R=0.39) | |

**Supplemental Online Material** for Ma & Ellison (2017) A new dominance concept and its application to diversity-stability analysis.

**Supplement to Section 3.1**: Computation and interpretation of new dominance metrics as well as comparisons between dominance metrics and diversity indices

> **Math Proof-**I: Mathematical proof of the functional (linear) relationship between the new *mean-crowding*-based community dominance index and Simpson's diversity index.
>
> Table S1. Fitting linear regression models ($D_c = a + bD$) between new community dominance index ($D_c$) and the existing diversity index ($D$)

**Supplement to Section 3.2**: Phenomenological modeling of community dominance and stability including Suppl. Tables (2)-(6), and Suppl. Figs (1)-(5).

> **Math Proof-II**: Criterion (parameter combinations) for determining community stability.

## Supplement to Section 3.1:

**Math Proof**-I: The linear relationship between the new mean-crowding-based *community dominance index* and Simpson's diversity index.

**Simpson diversity index** is defined as:

$$D = \sum_{i=1}^{n} p_i^2,$$

where n is the number of species and $p_i^2$ is the proportion of individuals belonging to the $i^{th}$ species,

namely $p_i = \dfrac{m_{s_i}}{\sum m_{s_i}}$.

**Community dominance index** is defined as:

$$D_c = \frac{m_c^*}{m_c}$$

where $m_c$ is the *mean population size* (abundance) per species, $m_c = \dfrac{\sum m_{s_i}}{n}$, $m_c^*$ is the *mean crowding* and is defined as, $m_c^* = m_c + V/m_c - 1$, $V$ is the variance of population size, *i.e.*, mean square deviation of $n$ species and is equal to: $V = \dfrac{\sum (m_{s_i} - m_c)^2}{n}$.

From the above definitions for Simpson's index ($D$) and our new community dominance index ($D_c$), we can derive the following linear function relationship $D_c$ and $D$.

$$D_c = \frac{m_c^*}{m_c} = \frac{m_c + V/m_c - 1}{m_c} = 1 + \frac{V}{m_c^2} - \frac{1}{m_c} = 1 + \frac{1}{n}\sum (m_c - m_{s_i})^2 \times \frac{n^2}{(\sum m_{s_i})^2} - \frac{n}{\sum m_{s_i}}$$

$$= 1 + n\sum (\frac{\frac{1}{n}\sum m_{s_i}}{\sum m_{s_i}} - \frac{m_{s_i}}{\sum m_{s_i}})^2 - \frac{n}{\sum m_{s_i}} = 1 + n\sum (\frac{1}{n} - p_i)^2 - \frac{n}{\sum m_{s_i}}$$

$$= 1 + n\sum (\frac{1}{n^2} - 2\frac{p_i}{n} + p_i^2) - \frac{n}{\sum m_{s_i}} = 1 + 1 - 2 + n\sum p_i^2 - \frac{n}{\sum m_{s_i}}$$

$$= n\sum p_i^2 - \frac{n}{\sum m_{s_i}}$$

$$= n * D - \frac{n}{\sum m_{s_i}}$$

Obviously, for a community sample, once observation is made, both $n$ and $\sum m_{s_i}$ are fixed (constant), therefore the above equation is a linear function between $D_c$ and $D$. Proof completed!

**Table S1**. Fitting linear regression models ($D_c = a + bD$) between new community dominance index ($D_c$) and the existing diversity index ($D$)*

| Subject ID | Berger-Parker Dominance Index | | | Shannon Entropy Index | | | Shannon Evenness Index | | | Simpson's Index | | | Simpson Community Evenness | | | n |
|---|---|---|---|---|---|---|---|---|---|---|---|---|---|---|---|---|
| | b | a | R | b | a | R | b | a | R | b | a | R | b | a | R | |
| 400 | 76.233 | -18.908 | 0.992 | -25.440 | 59.400 | 0.974 | -81.004 | 63.430 | 0.978 | **60.006** | **-0.028** | 1 | -671.578 | 59.806 | 0.723 | 29 |
| 401 | 91.551 | -11.698 | 0.962 | -36.585 | 85.473 | 0.954 | -147.117 | 100.248 | 0.961 | **93.002** | **-0.038** | 1 | -279.768 | 55.277 | 0.164 | 30 |
| 402 | 105.93 | -27.194 | 0.995 | -30.020 | 79.080 | 0.967 | -143.205 | 98.565 | 0.975 | **85.998** | **-0.039** | 1 | 199.425 | 49.042 | 0.126 | 31 |
| 403 | 45.096 | -13.625 | 0.991 | -17.264 | 32.560 | 0.988 | -47.396 | 33.672 | 0.977 | **32.005** | **-0.017** | 1 | -390.680 | 33.138 | 0.906 | 32 |
| 404 | 88.942 | -11.595 | 0.980 | -30.165 | 79.310 | 0.988 | -116.787 | 87.177 | 0.977 | **89.018** | **-0.046** | 1 | -1108.640 | 66.596 | 0.514 | 30 |
| 405 | 86.276 | -37.419 | 0.999 | -20.581 | 49.321 | 0.997 | -61.358 | 50.255 | 0.990 | **49.020** | **-0.037** | 1 | -742.236 | 50.375 | 0.890 | 31 |
| 406 | 101.939 | -27.238 | 0.985 | -33.374 | 79.568 | 0.989 | -121.524 | 85.435 | 0.991 | **79.004** | **-0.032** | 1 | -1278.120 | 73.492 | 0.691 | 31 |
| 407 | 126.464 | -20.967 | 0.986 | -46.216 | 113.350 | 0.985 | -164.807 | 117.505 | 0.995 | **113.007** | **-0.050** | 1 | -2700.970 | 110.392 | 0.870 | 28 |
| 408 | 45.202 | -3.736 | 0.955 | -16.222 | 42.731 | 0.932 | -52.755 | 45.844 | 0.941 | **54.993** | **-0.021** | 1 | -310.520 | 28.795 | 0.465 | 29 |
| 410 | 87.898 | -20.837 | 0.986 | -33.076 | 73.821 | 0.988 | -107.221 | 75.595 | 0.989 | **71.009** | **-0.032** | 1 | -1576.460 | 72.709 | 0.953 | 29 |
| 411 | 90.925 | -16.604 | 0.971 | -30.133 | 80.525 | 0.967 | -121.508 | 89.539 | 0.986 | **85.995** | **-0.034** | 1 | -1112.780 | 61.997 | 0.461 | 29 |
| 412 | 41.273 | -6.160 | 0.988 | -15.732 | 36.273 | 0.996 | -43.209 | 37.767 | 0.975 | **36.011** | **-0.024** | 1 | -403.415 | 37.735 | 0.816 | 28 |
| 413 | 51.119 | -9.034 | 0.972 | -19.554 | 45.708 | 0.971 | -62.781 | 49.131 | 0.982 | **46.975** | **-0.044** | 1 | -477.553 | 38.569 | 0.673 | 28 |
| 414 | 41.155 | -2.752 | 0.946 | -17.408 | 43.588 | 0.937 | -53.634 | 45.259 | 0.910 | **53.007** | **-0.052** | 1 | -309.402 | 27.896 | 0.458 | 32 |
| 415 | 66.014 | -12.815 | 0.991 | -29.132 | 57.713 | 0.975 | -84.688 | 60.325 | 0.978 | **56.001** | **-0.021** | 1 | -613.610 | 55.160 | 0.836 | 30 |
| 416 | 82.707 | -23.967 | 0.992 | -24.436 | 61.170 | 0.985 | -81.815 | 65.907 | 0.972 | **63.009** | **-0.073** | 1 | -1054.220 | 66.260 | 0.694 | 28 |
| 418 | 93.642 | -25.305 | 0.984 | -31.806 | 70.959 | 0.961 | -100.539 | 74.040 | 0.990 | **70.003** | **-0.059** | 1 | -706.122 | 68.852 | 0.801 | 31 |
| 420 | 74.177 | -13.356 | 0.990 | -25.985 | 64.102 | 0.996 | -90.578 | 68.158 | 0.992 | **63.987** | **-0.016** | 1 | -1440.980 | 70.750 | 0.808 | 28 |
| 423 | 115.039 | -24.352 | 0.975 | -43.142 | 96.685 | 0.978 | -147.915 | 99.420 | 0.989 | **95.999** | **-0.019** | 1 | -2106.650 | 96.310 | 0.901 | 29 |
| 424 | 134.283 | -26.185 | 0.978 | -47.797 | 119.388 | 0.983 | -186.818 | 128.044 | 0.981 | **122.999** | **-0.030** | 1 | -2932.300 | 118.000 | 0.642 | 28 |
| 429 | 101.616 | -29.516 | 0.984 | -37.538 | 75.766 | 0.959 | -120.077 | 77.469 | 0.984 | **74.008** | **-0.022** | 1 | -1093.780 | 72.869 | 0.896 | 29 |
| 430 | 51.449 | -0.690 | 0.951 | -17.191 | 51.893 | 0.967 | -62.370 | 52.830 | 0.921 | **83.996** | **-0.023** | 1 | -163.190 | 16.870 | 0.099 | 27 |
| 431 | 54.659 | -7.879 | 0.970 | -19.705 | 49.797 | 0.960 | -77.627 | 58.827 | 0.947 | **57.010** | **-0.028** | 1 | -33.711 | 21.474 | 0.028 | 27 |
| 432 | 110.308 | -25.278 | 0.982 | -38.914 | 87.777 | 0.954 | -138.186 | 91.979 | 0.981 | **87.994** | **-0.028** | 1 | -1433.800 | 86.099 | 0.806 | 29 |
| 435 | 71.670 | -14.998 | 0.983 | -28.762 | 63.581 | 0.976 | -94.911 | 66.895 | 0.982 | **62.010** | **-0.034** | 1 | -749.348 | 54.482 | 0.695 | 25 |
| 436 | 137.747 | -13.256 | 0.963 | -33.414 | 116.726 | 0.951 | -169.233 | 132.739 | 0.960 | **179.029** | **-0.097** | 1 | 300.760 | 32.450 | 0.042 | 32 |
| 437 | 139.841 | -50.273 | 0.997 | -36.635 | 91.695 | 0.989 | -130.372 | 95.590 | 0.992 | **91.021** | **-0.064** | 1 | -1718.690 | 93.807 | 0.769 | 33 |
| 439 | 40.120 | -5.985 | 0.975 | -18.131 | 42.182 | 0.978 | -60.033 | 46.281 | 0.970 | **41.997** | **-0.017** | 1 | -451.171 | 35.335 | 0.638 | 30 |
| 443 | 120.336 | -16.510 | 0.986 | -45.340 | 109.054 | 0.994 | -149.373 | 110.445 | 0.990 | **109.012** | **-0.049** | 1 | -2054.800 | 98.181 | 0.845 | 28 |
| 444 | 94.360 | -12.424 | 0.957 | -30.421 | 86.826 | 0.948 | -124.115 | 94.970 | 0.969 | **107.012** | **-0.034** | 1 | -1731.440 | 65.230 | 0.546 | 27 |
| 445 | 42.717 | -8.172 | 0.981 | -18.122 | 39.074 | 0.966 | -51.473 | 42.204 | 0.964 | **40.007** | **-0.026** | 1 | -290.157 | 32.194 | 0.664 | 30 |
| 446 | 81.504 | -25.828 | 0.992 | -30.098 | 60.290 | 0.974 | -82.816 | 63.401 | 0.752 | **57.955** | **0.012** | 1 | -107.987 | 47.465 | 0.145 | 29 |
| **Mean** | 84.131 | -17.642 | 0.979 | -29.011 | 70.168 | 0.972 | -102.414 | 75.28 | 0.967 | 75.378 | -0.035 | 1 | -923.247 | 59.3 | 0.611 | 29 |

*Regarding Berger-Parker and Shannon indexes, all the fitted linear models but two (2 out of 150 models) have the correlation coefficient $R > 0.95$ (the other two are 0.941, 0.752, but still exceed the critical $R=0.59$, at significance level $\alpha=0.001$, the degree of freedom=25). The Simpson's diversity index has perfect (analytically proved in Math Proof-I in Supplementary Documents) correlation with the new dominance index. The only less impressive correlations are with Simpson's evenness index but still very significant (average $R=0.61$, and median of $R=0.69$, exceed the critical value of $R=0.59$ for significance level $\alpha=0.001$, the degree of freedom=25). We argue that *Simpson's diversity index* is already a good measurement of *community evenness* because it measures the probability that any two randomly sampled individuals belong to the same species. *Simpson evenness index* is simply the diversity '*distributed*' over the species richness ($D/S$), and it may be a flawed metric for community evenness. The flaw may be due to the implicit assumption that all species are equivalent or at least similar with each other in terms of their *abundance distributions*.

## Supplement to Section 3.2: Phenomenological modeling of community dominance-stability dynamics

### Supplement to Modeling Methods

Note that there is not a commonly accepted definition for community stability in the literature. For example, Grimm & Wissel (1997) catalogued over 100 stability definitions, but few of them are quantitative, and none of them seem applicable for our objectives. The previously defined dominance indexes for both species and community scales set a necessary foundation for analyzing *community dominance*, but we still need some quantitative definitions and models to complete the analysis. We adopt a phenomenological, and exploratory approach to build mathematical models for community dominance dynamics.

We first define *community dominance stability* (shortened as *community stability* here after) with the following equation:

$$S_c(t) = \frac{D_c(t+1) - D_c(t)}{D_c(t)}. \qquad (1)$$

This definition of community stability essentially measures the change rate of community dominance; or equivalently, it measures the change rate of community diversity or composition.

Similarly, we may define population dominance stability (shortened as population stability) as the change rate of species dominance over time,

$$S_s(t) = \frac{D_s(t+1) - D_s(t)}{D_s(t)}, \qquad (2)$$

and there is a separate stability metric for each species.

We can assume that the dynamics of dominance is governed by a differential equation such as:

$$S(t) = \frac{dD(t)}{D(t)dt} = f[D(t), Z] \qquad (3)$$

where $D(t)$ is the dominance at time $t$, $Z$ is the vector of metadata factors, but it is optional, $S(t)$ is the stability. Eq.(3) can be applied at both community and species scales, depending on whether $D_c$ (for community) or $D_s$ (for species) is used.

Since we do not know the categorical form of function $f$, the modeling strategy we take is trial-and-error and also data-driven. Obviously, to perform data-driven modeling, the difference equation counterpart of the above differential equation should be preferred. For example, if we take Ricker model (see Cushing et al. 2001, Kot 2001), it has the differential equation:

$$S(t) = \frac{dD(t)}{D(t)dt} = r\left(1 - \frac{D(t)}{K}\right) \qquad (4)$$

Its discrete counterpart is:

$$D(t+1) = D(T)\exp[S(t)] \qquad (5)$$

where

$$S(t) = r[1 - D(t)/K] \qquad (6)$$

and $K$ is the maximum dominance value, similar to the *carrying capacity* in classical logistic model. Eq.(5) is the well-known logistic model.

Because we do not know the underlying mechanisms that generate any of the differential or difference equations, the above models are phenomenological. Through trial-and-error and pre-experiment exploratory curve-fittings, we found that the following five models are promising in describing the community dominance dynamics. These models are:

The logistic model for stability $S_c$ is:

$$S_c(t) = \frac{K}{1 + a * \exp[-rD_c(t)]} \tag{7}$$

where $K$, $a$, and $r$ are three parameters.

Logistic-Sine model:

$$S_c(t) = \frac{K}{1 + a * \exp[-rD_c(t)]} \sin\left(\frac{D_c(t)}{\pi}\right), \tag{8}$$

which is the product of logistic model [Eq. (11)] and a *Sine* function, and implies that there is a periodic factor that regulates the community stability in a multiplicative manner.

Linear model

$$S_c(t) = a + bD_c(t) \tag{9}$$

Linear-quadratic model (L-Q model)

$$S_c(t) = a + bD_c(t) + cD_c^2(t) + [D_c(t) - d]Sign(D_c(t) - d)[c(D_c(t) + d) + e)] \tag{10}$$

In this model, parameter $d$ is the interconnection point (*i.e.*, when *dominance* $D_c=d$) between the linear portion and quadratic portion; parameter $b_1=b-e$ determines the slope of the linear section, and parameter $c_2=2c$ determines the opening direction of parabola. These two parameters are particularly useful for assessing the types of the dominance-dependence relationships with the L-Q model. Detailed derivation of the parameters $b_1$ and $c_2$ are described in Math-Proof-II (derivations) in Supplementary Documents.

Quadratic-quadratic model (Q-Q model)

$$S_c = a + bD_c + cD_c^2 + (D_c - d)Sign(D_c - d)[e(D_c + d) + f)] \tag{11}$$

Similar with L-Q model, parameter $d$ is the interconnection point (*i.e.*, when *dominance* $D_c=d$) between the two parabolae; parameter $c_1=c-e$, and $c_2=c+e$ determine the opening direction of the first and second parabola, respectively. The three parameters are particularly useful for assessing the types of the dominance-dependence relationships with the Q-Q model. Again detailed derivation of the parameters $c_1$ and $c_2$ are described in Math-Proof-II (derivations) in Supplementary Documents.

For each of the stability models listed above, there is a corresponding dominance dynamics model, which can be derived from Eq.(1), *i.e.*,

$$D_c(t+1) = D_c(t)[1 + S_c(t)] \tag{12}$$

For example, with the logistic stability model [Eq.(7)], the corresponding model for dominance dynamics can be derived by substituting Eq.(7) into the above Eq. (12):

$$D_c(t+1) = D_c(t)\{1 + K/[1 + a\exp(-rD_c(t))]\}, \tag{13}$$

Similar to the *density-dependence* theory for population regulation from population ecology (*e.g.*, Berryman 1998, Kot 2001, Pastor 2008), the generalized stability model [Eq.(1)-(3)] may display three types of properties (behavior) locally: (*i*) *dominance-dependent stability* (DDS), *i.e.*, stability increases with dominance, (*ii*) *dominance-inversely-dependent stability* (DID)*, i.e.,* stability decreases with dominance, and (*iii*) *dominance-independent stability* (DIS)*,* stability does not change with dominance level. That is,

$$S(t) \propto kD(t) \qquad\qquad (14)$$

with $k < 0$, $k > 0$, $k = 0$, correspond to (*i*), (*ii*) and (*iii*), respectively. It is somewhat counterintuitive that $k < 0$ corresponds to high stability because we measure stability in terms of the magnitude of dominance change with Eq. (1)-(3) (actually instability).

In practice, except for the simple linear model case [Eq.(9), where parameter *b* is equivalent to *k* in Eq. (14)], we may not be able to determine the value of *k* since it may be infeasible to describe the nonlinear relationship with a single parameter such as *k*. However, in our case, the above selected nonlinear models are still simple enough that we can judge the dominance-stability dependence relationship piece-wisely, based on the combination of multiple parameters.

Specifically, the traditional logistic and linear model [Eq.(7) & (9)] may only capture one of the three *dominance-dependence* behaviors in a specific model, but the other three models may capture all three behaviors in a single model and therefore are much more flexible. The logistic-sine model can capture periodic fluctuation of the three types of dependence relationships.

## Supplement to Results:

Suppl. Tables (2-6) listed the model-fitting results including model parameters, the asymptotic standard errors of the parameters, and the adjusted determination coefficient ($R^2$). For the simple linear model [Eq. (9), Suppl. Table 4], we listed the results for all 57 subjects (32-healthy and 25-mixed cohort), even if model fitting failed. In fact, all the linear models except for one (the linear model for #s130) are statistically significant as discussed below, but linear models are not necessarily the most appropriate models for all the subjects. As a side note, in the tables for logistic model [Eq. (7), Suppl. Table (2)], logistic-sine [Eq. (8), Suppl. Table (3)], and linear model [Eq. (9), Suppl. Table (4)] we listed sample size (*n*), and this redundant information is omitted in the tables for linear-quadratic (L-Q) [Suppl. Table (5)] and quadratic-quadratic (Q-Q) [Suppl. Table (6)] models to save table space. But obviously, the sample size (*n*) information for L-Q and Q-Q models are the same as those for the other three phenomenological models.

When more than one model fit to the same subject's community successfully (statistically significant), we prefer models with fewer parameters, that is, in the order of linear, logistic, logistic-sine, linear-quadratic, and quadratic-quadratic models, with 2, 3, 5, and 6 parameters respectively. However, we realize that linear model may be too simple to capture the full spectrum of the dominance-stability relationships. Therefore, we treat linear model as a 'backup' model and assign priorities in the decreasing order of logistic, logistic-sine, linear-quadratic, quadratic-quadratic, and linear model.

Synthesizing the information from Suppl. Tables (2)–(6), we draw the following preliminary conclusions regarding the model selection from the five candidate models:

(*i*) The *logistic model* [Eq. (7), Suppl. Table (2), Suppl. Figures 1 & 2], when fitted to the data sets of HVMCs, is actually an *inverse* logistic curve. The first property that this inverse logistic curve displays is the *dominance-dependent stability* (DDS), which implies that the higher the dominance is, the higher the stability is. As a reminder, the higher stability is measured by the lower change rate of community dominance $S_c(t)$.

The parameter $K$ of the logistic curve exhibits the maximum change rate of dominance, or the maximum *instability* the community may exhibit. For example, in the case of *Subject*#412, showing in Figure 1, $K$=4.168, which suggests that the dominance may jump or crash as much as approximately 4 times, which can be computed from Eq. (12), $D_c(t+1) = D_c(t)[1 + S_c(t)] = 4.168 D_c(t)$. The parameter $r$ reflects the increase or decrease rate of community stability, it is largely equivalent to the $k$ in Eq. (14), *i.e.*, $S(t) \propto kD(t)$; the negative value of $r$ indicates that community stability increase with the growth of dominance—*dominance-dependent stability*, because negative $r$ signals the decrease of dominance change rate $S(t)$ with the increase of dominance level. The parameter $a$ is largely related to sampling scheme and may be influenced by factors such as starting date of sampling, similar to the start population size in traditional logistic population model. Its biological implication is not easy to identify, but may be ignored given that our objective is to reveal the change patterns and mechanisms in the dominance-stability paradigm, rather than to quantitatively predict the dynamics of dominance. The latter in our opinion is hardly feasible at this stage of human microbiome research, not to mention the experience from macro ecology, where community dynamics is often unpredictable.

The logistic model has its advantages over the other four models: First, it has the second least number of parameters (3 parameters, only next to the 2-parameter linear model), which makes the model fitting more robust than fitting the 5-parameter linear-quadratic model or the 6-parameter quadratic-quadratic model. Second, the parameter $K$ and $r$ possess clear biological meanings: $K$ is the asymptotic limit of dominance change rate and $r$ is the change rate. Therefore, the $K$ of logistic models listed in Suppl. Table 2 suggests that, for those subjects listed in Suppl. Table 2, the stability or the change rate of community dominance is bounded, and the variation among individual subjects is in the range between 0.188 and 23.194. The corresponding $r$ of those logistic models in Suppl. Table 2 indicates that the stability is density-dependent, and the variation of change rate among individual subjects is between 0.126 and 1.813.

However, the logistic model has its limitation. The model fits to just slightly more than one-fourth (15 out of 57) of the subjects well [Suppl. Table (2)], judged from $R^2$. The failures in fitting logistic model fall into the following three categories: *too small $R^2$* (smaller than 0.30), *too large* standard error of the model parameter, and *erratic* parameter values. For example, with subject#408, $R^2$ =0.78 is not small at all, but the standard error of parameter $K$ is rather large ($K$=22.644, Standard Error=2602.638), and we deem the fitting of logistic model to this subject as a failure. Similarly, the standard errors of parameters $K$ & $a$ of subject #411 are too large to be reasonable. When the standard error of a parameter is too large, the model-fitting becomes instable and the parameter estimation may become unreliable. In the case of subject#s52, the value of parameter $K$ (=15152070) is simply too big to make sense biologically.

Although there is no absolute criterion for determining the sufficiently large $R^2$ for non-linear models, too small $R^2$ is certainly unacceptable because, like the other two scenarios, the model becomes unreliable when any of the three issues identified above occurs. In Suppl. Table (2), we include the parameters for three subjects (#408, #411, and #s52) to illustrate the three types of failures in model fitting. They are not counted as successful modeling fittings and not used in devising the summary table [Table (2)] in the main text of the paper. The three issues (associated with the three criteria, $R^2$, *sanity* of the parameter and its standard error) described above are also responsible for the failures of fitting the other four models to the datasets. In the case of Logistic model, we listed the failed cases in the results table [Table (2)] to demonstrate the criteria we adopted to judge the success/failure of model fitting. With the other four models, we do not list the failed models in their respective tables [Table (3)-(6)].

Except for the case of linear models [Suppl. Table (4)], there is not any absolute criterion for judging the success or failure of fitting to the nonlinear models. Nevertheless, since the results from fitting to the linear model demonstrated that all but one (out of totally 57 subjects) are statistically significant, there is a reasonable expectation that nonlinear models such as logistic model, L-Q, and Q-Q models, especially the last two, should outperform the simple linear model. In fact, linear model is a special case of both L-Q, and Q-Q models, and logistic model may also contain a linear section. Therefore, if the $R^2 (R)$ is significant (sufficiently large) for linear model, it should also be significant for the nonlinear models. In other words, $R^2 (R)$ should not be a big concern in judging the fitness of the nonlinear models in our study. Instead, model *over-fitting* that could lead to the *erratic* parameters (*e.g.*, #s52) or standard error (#408), the two issues discussed above, must be scrutinized carefully for their biological rationality (sanity).

As argued previously, while statistical significance is a necessary condition for selecting a model, biological significance should be a sufficient condition. To determine the biological significance, we need to compare the five different models—logistic, logistic-sine, linear, L-Q, and Q-Q models. Except for subject#408, #411, and #s52, Suppl. Table (2) lists the candidate logistic models that passed our statistical validity checking for the logistic model, and whether or not they are appropriate for a specific subject will be determined by the comparisons with the other four models. As explained and displayed in Table (2) of the main document, ultimately, only 6 out of 15 logistic models (that passed statistical validity checking) managed to pass our testing of their biological significance. Those 6 models are appropriate for the following 6 subjects: #400, #412, #439, #s40, #s59, and #s96, respectively. Suppl. Figure (1) and (2) are the graphs of the logistic model for subject#412 and #s40, respectively.

Table S2. The logistic model for describing the *stability* of *community dominance**

| Subject ID | $K$ | Std. Err. of $K$ | $r$ | Std. Err. of $r$ | $a$ | Std. Err of $a$ | $R^2$ | $n$ |
|---|---|---|---|---|---|---|---|---|
| 400 | 4.741 | 1.540 | -0.206 | 0.059 | 0.026 | 0.0506 | 0.91 | 28 |
| 408 | 22.644 | 2602.6 | -1.813 | 2.562 | 0.00000 | 0.0022 | 0.78 | 28 |
| 411 | 23.194 | 278.1 | -0.182 | 0.151 | 0.712 | 11.001 | 0.62 | 28 |
| 412 | 4.168 | 1.010 | -0.647 | 9.893 | 0.00002 | 0.0056 | 0.99 | 27 |
| 415 | 0.801 | 0.227 | -0.346 | 0.157 | 0.00001 | 0.0001 | 0.53 | 29 |
| 416 | 0.937 | 0.186 | -0.411 | 0.040 | 0.00000 | 0.00001 | 0.63 | 27 |
| 418 | 1.402 | 0.396 | -0.386 | 0.022 | 0.00000 | 0.00001 | 0.80 | 30 |
| 420 | 2.108 | 0.333 | -0.604 | 0.064 | -0.00001 | 0.00001 | 0.71 | 27 |
| 423 | 2.133 | 0.351 | -0.191 | 0.088 | 0.00074 | 0.0028 | 0.73 | 28 |
| 424 | 1.175 | 0.459 | -0.126 | 0.103 | 0.00364 | 0.0206 | 0.44 | 27 |
| 436 | 1.478 | 0.509 | -0.632 | 0.162 | 0.00000 | 0.00003 | 0.46 | 31 |
| 439 | 3.277 | 0.465 | -1.029 | 1.069 | 0.00001 | 0.0003 | 0.87 | 29 |
| 445 | 0.872 | 0.259 | -0.724 | 0.237 | 0.00001 | 0.00008 | 0.50 | 29 |
| *Mean* | 2.099 | 0.521 | -0.482 | 1.081 | 0.003 | 0.007 | 0.69 | 28 |

* The two models (for #408, #411 highlighted in red color) failed to pass our three statistical criteria, but listed here to demonstrate the failure causes. Among the 13 models that passed our statistical validity checking, only 3 (highlighted in green color) are ultimately selected after passing biological interpretations.

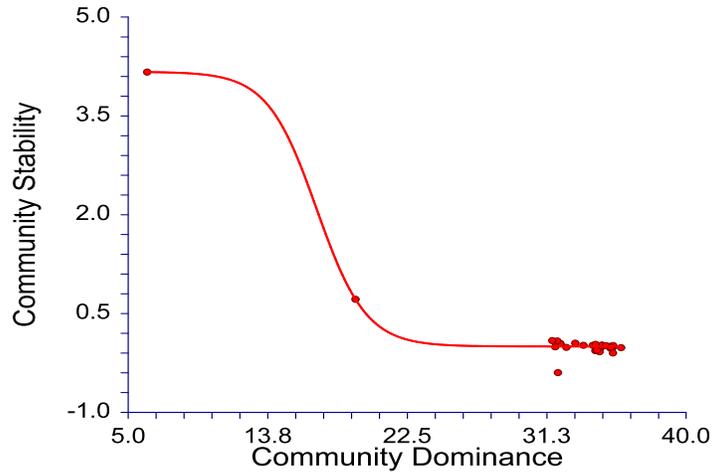

Figure S1. Logistic stability model for Subject #412 (Healthy)

We noticed that some of the failures in fitting logistic model are due to the extensive fluctuations of dominance dynamics (change rates). To find a possible remedy, we introduced the *logistic-sine* model [Eq. (8)], which is an extension of the standard logistic model, as discussed below.

(*ii*) The *logistic-sine* function is a multiplicative modification to the standard 3-parameter logistic model with a *sine* function $[\sin(D_c(t)/\pi]$. We also tried additive modification, but the model fitting performed not as well as the multiplicative version. Obviously, the *sine* function was applied with the hope to capture the possible periodic change of microbial community due to woman's menses cycle. Somewhat surprisingly, this modified logistic function seems to have limited suitability, and we only found four subjects (candidates: #405, #420, #431, #445) where this model seems to demonstrate its advantages [Suppl. Table (3)], and ultimately assigned the logistic-sine model as the primary model only to the 2 subjects (#420 and #431, highlighted in green in Table 3).

An interesting pattern that the logistic-sine model revealed is the attenuating fluctuation of community stability. The community displays significant attenuation with the increase of community dominance, but the magnitude of the attenuation as well as the period may differ among communities. Suppl. Figure (2) displays the graph of the logistic-sine model for *subject*#420, and the reason for sudden big crash is unknown.

Although the logistic-sine model achieved limited success with our datasets (2 out of 32 subjects), it may be very useful when longer longitudinal datasets (*e.g.,* one year at least, rather than three months with our datasets, are available for analysis). In general, a message this model signals is that community stability can be continuously fluctuating with three types *of dominance-stability dependences* [*dominance-dependent*, *dominance-inversely dependent*, and *dominance-independent* (only at some points)] repeating periodically.

Table S3. The logistic-sine model for describing the stability of community dominance

| Subject ID | $K$ | Std. Err. of $K$ | $r$ | Std. Err. of $r$ | $a$ | Std. Err. of $a$ | $R^2$ | $n$ |
|---|---|---|---|---|---|---|---|---|
| 405 | 0.008 | 0.145 | 0.022 | 0.407 | -2.067 | 24.909 | 0.60 | 30 |
| 420 | -0.294 | 0.209 | 0.048 | 0.036 | -1.677 | 0.674 | 0.71 | 27 |
| 431 | -0.142 | 0.154 | 0.060 | 0.067 | -1.822 | 1.264 | 0.36 | 26 |
| 445 | -0.025 | 0.010 | 0.009 | 0.00093 | -1.097 | 0.006 | 0.52 | 29 |
| *Mean* | -0.113 | 0.130 | 0.035 | 0.128 | -1.666 | 6.713 | 0.548 | 28 |

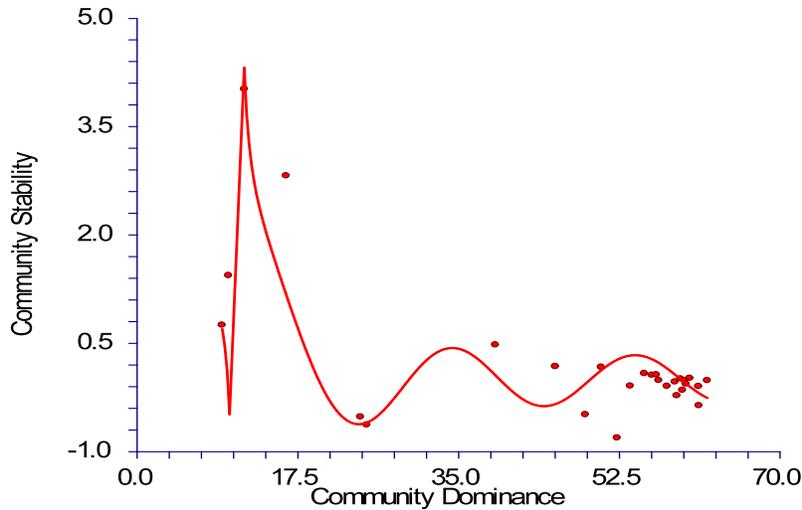

Figure S2. Logistic-Sine stability model for Subject #420 (Healthy)

(*iii*) *Linear model*, the simplest among the five models we conceived for describing community dominance dynamics, turned out to fit all 32 subjects successfully, judged from the linear correlation coefficient [Suppl. Table (4)]. Another important characteristic displayed by these linear models is that their regression *slopes* are negative (*b*<0). This suggests the potential *dominance-dependent stability* pattern. However, we err on the side of caution and do not rush to accept the conclusion for all subjects. This is because, although the linear statistical relationship is supported by the data, non-linear relationships revealed by the other four models may not be ignored (Suppl. Figs 1-2, Figs 4-5). Furthermore, the extensive debates on diversity-stability paradigm in ecological literature suggest that the dominance-stability relationship is less likely to be as straightforward as the monotonically dominance-dependent relationship suggested by the linear model. Therefore, we accept linear model as primary model for a subject only if the other four models are inferior to linear model. Suppl. Figure (3) is the graph of the linear model for subject#405.

Table S4. The *linear* model for describing the *stability of community dominance*

| Subject ID | *a* | Standard. Error of *a* | *b* | Standard Error of *b* | *R* | *n* |
|---|---|---|---|---|---|---|
| 400 | 2.974 | 0.407 | -0.061 | 0.009 | 0.81 | 28 |
| 401 | 0.594 | 0.180 | -0.012 | 0.004 | 0.53 | 29 |
| 402 | 1.087 | 0.225 | -0.019 | 0.004 | 0.67 | 30 |
| 403 | 1.361 | 0.188 | -0.047 | 0.007 | 0.80 | 31 |
| 404 | 0.488 | 0.218 | -0.010 | 0.005 | 0.37 | 29 |
| 405 | 1.551 | 0.182 | -0.033 | 0.004 | 0.85 | 30 |
| 406 | 0.800 | 0.233 | -0.014 | 0.004 | 0.52 | 30 |
| 407 | 1.783 | 0.451 | -0.021 | 0.006 | 0.56 | 27 |
| 408 | 1.382 | 0.395 | -0.074 | 0.022 | 0.54 | 28 |
| 410 | 0.642 | 0.224 | -0.012 | 0.004 | 0.49 | 28 |
| 411 | 1.704 | 0.346 | -0.037 | 0.008 | 0.67 | 28 |
| 412 | 4.150 | 0.345 | -0.123 | 0.010 | 0.92 | 27 |
| 413 | 0.789 | 0.265 | -0.028 | 0.011 | 0.45 | 27 |
| 414 | 0.941 | 0.245 | -0.056 | 0.015 | 0.57 | 31 |
| 415 | 1.465 | 0.240 | -0.032 | 0.005 | 0.75 | 29 |
| 416 | 1.451 | 0.240 | -0.032 | 0.006 | 0.76 | 27 |
| 418 | 1.739 | 0.242 | -0.028 | 0.004 | 0.80 | 30 |
| 420 | 1.973 | 0.441 | -0.037 | 0.009 | 0.64 | 27 |
| 423 | 1.559 | 0.278 | -0.020 | 0.004 | 0.73 | 28 |
| 424 | 1.260 | 0.256 | -0.016 | 0.003 | 0.68 | 27 |
| 429 | 0.979 | 0.215 | -0.015 | 0.003 | 0.66 | 28 |
| 430 | 0.539 | 0.206 | -0.037 | 0.015 | 0.45 | 26 |
| 431 | 0.932 | 0.320 | -0.037 | 0.004 | 0.46 | 26 |
| 432 | 1.269 | 0.299 | -0.017 | 0.004 | 0.63 | 28 |
| 435 | 1.769 | 0.355 | -0.041 | 0.009 | 0.71 | 24 |
| 436 | 1.409 | 0.317 | -0.031 | 0.008 | 0.59 | 31 |
| 437 | 1.220 | 0.262 | -0.015 | 0.003 | 0.64 | 32 |
| 439 | 1.775 | 0.389 | -0.078 | 0.018 | 0.64 | 29 |
| 443 | 0.718 | 0.241 | -0.008 | 0.004 | 0.43 | 27 |
| 444 | 0.486 | 0.221 | -0.014 | 0.007 | 0.39 | 26 |
| 445 | 1.537 | 0.220 | -0.073 | 0.011 | 0.79 | 29 |
| 446 | 0.644 | 0.184 | -0.014 | 0.004 | 0.55 | 28 |
| *Mean* | 1.343 | 0.276 | -0.034 | 0.008 | 0.627 | 28 |

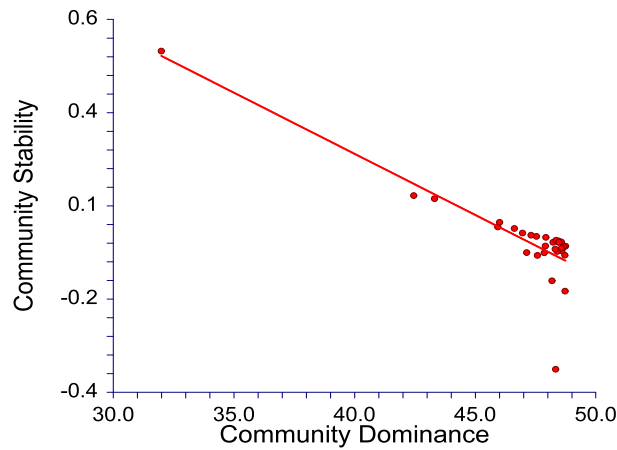

Figure S3. Linear stability model for subject #405 (R=0.84, Healthy)

(*iv*) The *linear-quadratic* (L-Q) model is *a piece-wise* integration of a *linear function* followed by a *quadratic function*. This is a 5-parameter model, *a*, *b*, *c*, *d*, *e* with *d* as the joint point of the two pieces. The value of *d* is the *community dominance* level at which linear curve and quadratic curve interconnect [*e.g.*, *d*=28.34 in Suppl. Figure (4) for the community of *subject*#402].

Suppl. Table (5) lists 15 candidate L-Q models from fitting to 32 subjects. Judging from $R^2$, the number of candidate communities selected for the L-Q model is the second largest group among the five models, only next to the linear model. In fact, the L-Q models perform better than the linear models if judged from $R^2$ [*R* is used in Suppl. Table (4) for linear model because *R* can be compared with significance level $R_0$]. In some cases, 6-parameter Q-Q models outperform 5-parameter L-Q models slightly judged from $R^2$, but we caution to select the former based on the value of $R^2$. Instead, based on the principle of parsimony, we select the Q-Q model only if the L-Q model and other models fail to describe stability dynamics adequately, as discussed below.

Besides parameter *d*, which is the joint point of the linear and quadratic curves, two parameter $b_1$=*b-e*, and $c_2$=2*c* are particularly important. Both parameter $b_1$ and $c_2$ are computed parameters, rather than being fitted directly, and they are listed as the last two columns in Suppl. Table (5). Parameter $b_1$ is the slope of linear piece of L-Q model and parameter $c_2$ determines the opening directions of parabola (opens up when $c_2$>0 or down when $c_2$<0). Specifically, parameter *d* determines when the relationship switches from linear to non-linear form. The combinations of $b_1$ & $c_2$ determine the types of stability-dominance dependence relationship. There can be four combinations with the values of parameters $b_1$ & $c_2$: $b_1>0$ corresponding to dominance-inversely-dependent stability during the linear phase, $b_1<0$ to dominance-dependent stability during the linear phase, $c_2>0$ corresponding to the open-up parabola, and $c_2<0$ to open-down parabola. During the nonlinear parabola phase, all three types of dominance-stability dependence relationships are possible, and there is a dominance-independent point (vertex of parabola) that acts as the *tipping point* between the other two types. The dominance-independent point (vertex or tipping point) is a stable equilibrium point when the parameter $c_2>0$, but unstable when $c_2<0$. Furthermore, the interconnection point between linear and quadratic curves can also be an equilibrium point. Obviously, if $b_1>0$, the equilibrium is unstable; if $b_1<0$, the equilibrium depends on the parameter $c_2$. Therefore, there are potentially two equilibrium points with this model: one occurs during the transition from linear to quadratic phase, and the other is the vertex of the parabola.

For example, Suppl. Fig. 4 shows the graph of L-Q model for *subject*#402 with both $b_1$ and $c_2$ being positive. The first inflection point (equilibrium) in the graph is determined by parameter $d$ and is unstable, but the possibly second one (vertex of the parabola) is stable, although the vertex of the parabola is not displayed in the graph due to the data range limitation. The linear phase represents the *dominance inversely dependent stability* mechanism and the quadratic phase represents the dominance dependent stability mechanism. Both phases are connected by an unstable equilibrium point corresponding to parameter *d=28.34*.

From the 15 candidate communities listed in Suppl. Table (5), we designate L-Q model as the primary model for selected 10 communities (highlighted in green in the following Suppl. Table 5, also see Table 2 in the main text), those are #402, #408, #411, #416, #423, #429, #435, #436, #437, and #445. They are selected only if they outperform more parsimonious (few parameters) logistic model, logistic-sine model, and linear model, judged by $R^2$, the standard errors of the parameters, and visual check of their model graphs, as well as the parameters' biological interpretations.

Table S5. The L-Q model for describing the stability of community dominance

| Subject ID | $a$ | Std Err of $a$ | $b$ | Std Err of $b$ | $c$ | Std Err of $c$ | $d$ | Std Err of $d$ | $e$ | Std Err of $e$ | $R^2$ | $b_1=-b-e$ | $c_2=-2c$ |
|---|---|---|---|---|---|---|---|---|---|---|---|---|---|
| 400 | 0.611 | 4.602 | 0.007 | 0.177 | -0.001 | 0.00169 | 34.380 | 4.767 | 0.182 | 0.177 | 0.93 | -0.175 | -0.002 |
| 402 | 0.331 | 0.676 | 0.014 | 0.027 | 0.00033 | 0.00017 | 28.341 | 1.690 | -0.108 | 0.027 | 0.68 | 0.122 | 0.0007 |
| 405 | 0.337 | 0.000 | 0.020 | 0.000 | -0.00057 | 0.00000 | 42.136 | 0.000 | 0.056 | 0.000 | 0.72 | -0.036 | -0.0011 |
| 408 | 49.023 | 21.883 | -5.947 | 2.738 | 0.00067 | 0.00087 | 8.187 | 0.095 | 5.862 | 2.738 | 0.85 | -11.8 | 0.0013 |
| 411 | 2.001 | 1.073 | -0.067 | 0.045 | -0.00027 | 0.00038 | 27.018 | 2.858 | 0.106 | 0.045 | 0.68 | -0.173 | -0.0005 |
| 412 | 0.587 | 7.999 | 0.004 | 0.476 | -0.00201 | 0.00707 | 23.613 | 4.075 | 0.268 | 0.476 | 0.99 | -0.264 | -0.004 |
| 416 | 2.210 | 0.794 | -0.059 | 0.040 | 0.00099 | 0.00035 | 23.740 | 2.122 | -0.149 | 0.040 | 0.74 | 0.09 | 0.00198 |
| 418 | 1.758 | 3.113 | -0.043 | 0.101 | -0.00000 | 0.00082 | 45.896 | 4.469 | 0.051 | 0.101 | 0.83 | -0.094 | -0.00000 |
| 423 | -0.371 | 3.137 | 0.020 | 0.081 | -0.00029 | 0.00051 | 61.448 | 4.463 | 0.071 | 0.081 | 0.77 | -0.051 | -0.00058 |
| 429 | -28.13 | 8.567 | 0.728 | 0.225 | 0.00058 | 0.00037 | 38.752 | 0.251 | -0.888 | 0.225 | 0.67 | 1.616 | 0.0011 |
| 435 | 1.714 | 2.694 | -0.054 | 0.115 | -0.00042 | 0.00119 | 28.569 | 4.516 | 0.119 | 0.114 | 0.73 | -0.173 | -0.00084 |
| 436 | -5.741 | 8.829 | 0.171 | 0.255 | -0.00142 | 0.00178 | 51.078 | 6.640 | 0.223 | 0.255 | 0.47 | -0.052 | -0.0028 |
| 437 | -4.670 | 7.475 | 0.127 | 0.183 | -0.00094 | 0.00112 | 66.958 | 5.260 | 0.180 | 0.183 | 0.54 | -0.053 | -0.00188 |
| 439 | 2.135 | 0.657 | -0.136 | 0.059 | -0.00175 | 0.00118 | 14.192 | 0.610 | 0.284 | 0.059 | 0.89 | -0.42 | -0.0035 |
| 445 | 1.608 | 3.139 | -0.091 | 0.246 | 0.00032 | 0.00473 | 19.639 | 5.926 | 0.016 | 0.246 | 0.66 | -0.107 | 0.00064 |
| *Mean* | 1.560 | 4.976 | -0.354 | 0.318 | 0.000 | 0.001 | 34.263 | 3.183 | 0.418 | 0.318 | 0.74 | -0.772 | -0.00077 |

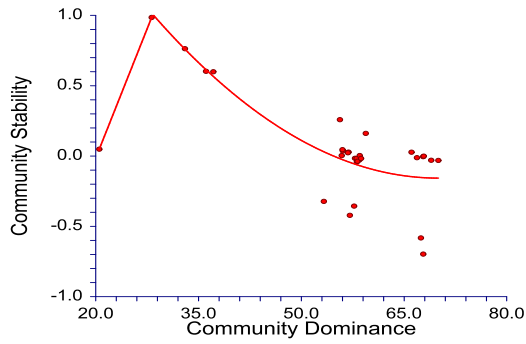

Figure S4. Linear-Quadratic stability model for Subject #402 (Healthy)

(*v*) The *quadratic-quadratic* (Q-Q) model is a piecewise interconnection of two quadratic curves with six parameters, *a-f*. The model is similar to the previous *linear-quadratic* (L-Q) model, but both pieces are quadratic functions, rather than being the combination of linear and quadratic function. This difference makes Q-Q model more flexible in fitting to data because it removes the limitation that the stability-dominance relationship is linear in the initial stage, and allows more flexible non-linear relationships across the whole spectrum of the relation. Nevertheless, the increase of the number of parameters in Q-Q models also raises the requirements for the data sets, and also raise the risk of over-fitting, because theoretically one may always find polynomial fitting to a set of data with sufficiently high order, not to mention piece-wisely connected polynomials. Therefore, considering the Occam's razor rule, we select the Q-Q model only if it significantly outperforms logistic, logistic-sine, and L-Q models because the other models are more parsimonious in terms of the number of model parameters.

Given the similarity with L-Q model, our discussion of the Q-Q model is brief and focused on its difference with L-Q model, and one may easily infer most of the similar properties of the Q-Q model by referring to the L-Q model. Suppl. Table (6) lists 13 candidate Q-Q models from fitting to 32 subjects. From the 13 candidate communities listed in Suppl. Table (6), we designate Q-Q model as the primary model for selected 3 communities, those are #415, #418, #446; highlighted in green in Suppl. Table (6)]. They are selected because they outperform the other more parsimonious (few parameters) models according to $R^2$, the standard errors of the parameters, and visual check of their model graphs.

Suppl. Table 6. The quadratic-quadratic model for describing the *stability* of *community dominance*

| Subject ID | a | Std. Err of a | b | Std. Err of b | c | Std. Err of c | d | Std. Err of d | e | Std. Err of e | f | Std. Err of f | $R^2$ | $c_1$=c-e | $c_2$=c+e |
|---|---|---|---|---|---|---|---|---|---|---|---|---|---|---|---|
| 400 | 1.349 | 4.743 | -0.076 | 0.208 | 0.0009 | 0.003 | 37.266 | 12.068 | -0.0029 | 0.0030 | 0.264 | 0.208 | 0.93 | 0.004 | -0.002 |
| 402 | -6.679 | 5.385 | 0.373 | 0.196 | -0.0053 | 0.002 | 43.009 | 3.901 | 0.0036 | 0.0022 | -0.171 | 0.196 | 0.69 | -0.0089 | -0.0017 |
| 411 | 3.255 | 2.109 | -0.202 | 0.203 | 0.0031 | 0.005 | 29.469 | 18.902 | -0.0037 | 0.0051 | 0.239 | 0.213 | 0.69 | 0.0068 | -0.0006 |
| 412 | 0.317 | 0.000 | 0.062 | 0.000 | -0.0043 | 0.000 | 22.840 | 0.000 | 0.0003 | 0.000 | 0.210 | 0.000 | 0.99 | -0.0046 | -0.004 |
| 415 | -3.361 | 0.000 | 0.269 | 0.000 | -0.0007 | 0.000 | 18.353 | 0.000 | 0.0017 | 0.0000 | -0.385 | 0.000 | 0.73 | -0.0024 | 0.001 |
| 416 | -2.369 | 2.629 | 0.206 | 0.121 | -0.0038 | 0.002 | 37.773 | 2.064 | 0.0026 | 0.0016 | -0.105 | 0.121 | 0.77 | -0.0064 | -0.0012 |
| 418 | -1.945 | 3.459 | 0.176 | 0.144 | -0.0032 | 0.002 | 43.002 | 2.732 | 0.0032 | 0.0018 | -0.168 | 0.144 | 0.86 | -0.0064 | -0.00001 |
| 423 | -0.438 | 3.232 | 0.024 | 0.087 | -0.0004 | 0.001 | 60.610 | 6.412 | -0.0002 | 0.0007 | 0.067 | 0.087 | 0.77 | -0.0002 | -0.0006 |
| 432 | 5.206 | 2.600 | -0.195 | 0.115 | 0.0019 | 0.002 | 49.667 | 15.186 | -0.0012 | 0.0016 | 0.089 | 0.132 | 0.61 | 0.0031 | 0.0007 |
| 435 | 3.331 | 3.803 | -0.207 | 0.173 | 0.0031 | 0.003 | 35.138 | 50.902 | -0.0039 | 0.0030 | 0.262 | 0.207 | 0.73 | 0.007 | -0.0008 |
| 439 | 3.170 | 0.982 | -0.363 | 0.156 | 0.0099 | 0.007 | 15.263 | 1.652 | -0.0130 | 0.0072 | 0.492 | 0.159 | 0.89 | 0.0229 | -0.0031 |
| 445 | 0.683 | 2.967 | 0.015 | 0.281 | -0.0028 | 0.008 | 19.285 | 3.266 | 0.0023 | 0.0077 | -0.027 | 0.281 | 0.66 | -0.0051 | -0.0005 |
| 446 | -12.71 | 4.518 | 0.572 | 0.190 | -0.0066 | 0.002 | 43.061 | 0.882 | -0.0027 | 0.0020 | 0.333 | 0.190 | 0.59 | -0.0039 | -0.0093 |
| **Mean** | -0.450 | 3.295 | 0.032 | 0.180 | 0.000 | 0.003 | 34.412 | 8.760 | -0.001 | 0.003 | 0.054 | 0.187 | 0.76 | 0.0004 | -0.0017 |

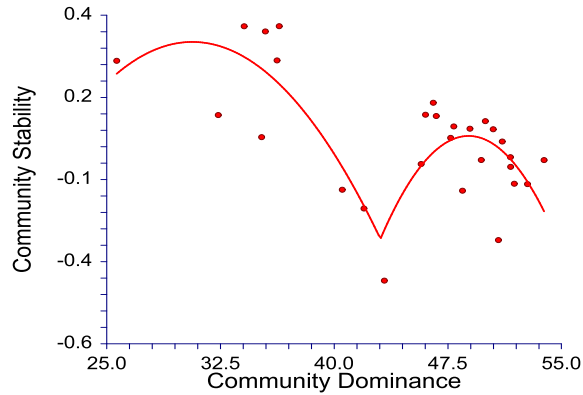

Figure S5. Quadratic-Quadratic stability model for Subject#446 (Healthy)

Suppl. Fig. (5) is the Q-Q model graph for *subject*#446, and it has three potential equilibrium points, where the first and third equilibriums should be unstable, and the second one (the joint point of two parabolas) is obviously stable. Similar to the L-Q model, parameters $c_1$ and $c_2$, which are the 2nd-order (squared) coefficients of the Q-Q model as explained in the main text, determine the opening directions of the parabolas, and hence the types of the dominance-stability relationships. In general, parameter *d,* which equals the dominance level at which the two parabolas connect, determines the transition (connection or joint) point between the two parabolas. The connection point may be stable or not, depending on the open directions of both parabolas, as well as the relative location of the connection point (joint). Parameter $c_1$ determines whether or not the first equilibrium (vertex of the first parabola) is stable, and $c_2$ determines the fate of the third equilibrium (vertex of the second parabola). When $c_1 < 0$, the first parabola opens down and the corresponding equilibrium should be unstable, and when $c_1 > 0$, it opens up and the equilibrium should be stable. The sign of parameter $c_2$ similarly determines the stability of the third equilibrium point. In Suppl. Fig. (5), both pieces of parabolas open down since the values of $c_1$ & $c_2$ are negative.

**Math Proof-II** (Derivations): Criterion (parameter combinations) for determining community stability with L-Q and Q-Q models.

For logistic, logistic-since, and linear models, the criteria for determining community stability are relatively simple and there is not a need to have detailed discussion here (one can get full information in the main text or above relevant paragraphs. With Linear-Quadratic (L-Q) model and Quadratic-Quadratic (Q-Q) model, we present the following mathematical derivations to support the discussion in main text.

L-Q model is in the form [Eq. (10) in main text]:

$$S_c(t) = a + bD_c(t) + cD_c^2(t) + [D_c(t) - d]Sign(D_c(t) - d)[c(D_c(t) + d) + e)]$$

It is the interconnection of the following linear and quadratic functions:

$$f(x) = a_1 + b_1 x \qquad x \le D$$
$$f(x) = a_2 + b_2 x + c_x x^2 \qquad x > D$$

According to NCSS (2007), the parameters of the above three functions have the following relationships:

| | | | |
|---|---|---|---|
| $a=(a_1+a_2)/2$ | $b=(b_1+b_2)/2$ | $c=c_2/2$ | $e=(b_2-b_1)/2$ |
| $a_1=a+2cd+de$ | $b_1=b-e$ | | |
| $a_2=a-2cd-de$ | $b_2=b+e$ | $c_2=2c$ | |

It is clear from the above relationships, parameter $d$ is the interconnection point (*i.e.*, when *dominance* $D_c=d$) between linear portion and quadratic portion; parameter $b_1=b-e$ determines the slope of the linear section, and parameter $c_2=2c$ determines the opening direction of parabola. When $c_2>0$, the parabola opens up, and it opens down when $c_2<0$. These three parameters ($d$, $b_1$, $c_2$) are particularly useful for assessing the types of the dominance-dependence relationships with the L-Q model.

Q-Q model is in the form [Eq. (11) in main text]:

$$S_c = a + bD_c + cD_c^2 + (D_c - d)Sign(D_c - d)[e(D_c + d) + f)]$$

It is the interconnection of the following two common quadratic functions:

$$f(x) = a_1 + b_1 x + c_1 x^2 \qquad x \le D$$
$$f(x) = a_2 + b_2 x + c_x x^2 \qquad x > D$$

According to NCSS (2007), the parameters of the above three functions have the following relationships:

| | | |
|---|---|---|
| $a=(a_1+a_2)/2$ | $b=(b_1+b_2)/2$ | $c=(c_1+c_2)/2$ |
| $e=(c_2-c_1)/2$ | $f=(b_2-b_1)/2$ | |
| $a_1=a-2ed+df$ | $b_1=b-f$ | $c_1=c-e$ |
| $a_2=a+2ed-df$ | $b_2=b+f$ | $c_2=c+e$ |

Similar with L-Q model, parameter $d$ is the interconnection point (*i.e.*, when *dominance* $D_c=d$) between the two parabolae. Parameter $c_1=c-e$, and $c_2=c+e$ determine the opening direction of the first and second parabola, respectively. When $c_1>0$, the first piece of parabola opens up, and it opens down when $c_1<0$, and similar rule applies to the second piece of parabola. The three parameters ($d$, $c_1$, $c_2$) are particularly useful for assessing the types of the dominance-dependence relationships with the Q-Q model.